\documentclass[a4paper,11pt]{article}

\usepackage{amsmath,amssymb,bm,color}
\usepackage{appendix}
\usepackage{graphicx}
\usepackage{xcolor}
\usepackage{cancel}
\usepackage{jcappub}
\usepackage[T1]{fontenc} 

\newcommand{\be}{\begin{equation}}
\newcommand{\ee}{\end{equation}}
\newcommand{\bea}{\begin{eqnarray}}
\newcommand{\eea}{\end{eqnarray}}

\title
{Baryogenesis, Primordial Black Holes and MHz-GHz Gravitational Waves}

\author[a]{Thomas C. Gehrman}
\author[b]{Barmak Shams Es Haghi}
\author[a]{Kuver Sinha}
\author[a,c]{Tao Xu}

\affiliation[a]{Department of Physics and Astronomy, University of Oklahoma, Norman, OK 73019, USA}
\affiliation[b]{Texas Center for Cosmology and Astroparticle Physics, Weinberg Institute, Department of Physics, The University of Texas at Austin, Austin, TX 78712, USA}
\affiliation[c]{Racah Institute of Physics, Hebrew University of Jerusalem, Jerusalem 91904, Israel}

\emailAdd{thomas.gehrman@ou.edu}
\emailAdd{shams@austin.utexas.edu}
\emailAdd{kuver.sinha@ou.edu}
\emailAdd{tao.xu@ou.edu}

\abstract{

Gravitational waves (GWs) in the MHz - GHz frequency range are motivated by a host of early Universe phenomena such as oscillons, preheating, and cosmic strings. We point out that baryogenesis too serves as a motivation to probe GWs in this frequency range. The connection is through primordial black holes (PBHs): on the one hand, PBHs induce baryogenesis by Hawking evaporating into a species that has baryon number and $CP$ violating decays; on the other, PBHs induce GWs through second order effects when the scalar fluctuations responsible for their formation re-enter the horizon. We describe the interplay of the parameters responsible for successful baryogenesis on the plane of the strain and frequency of the induced GWs, being careful to delineate regimes where PBH domination or washout effects occur. We provide semi-analytic scalings of the GW strain with the baryon number to entropy ratio and other parameters important for baryogenesis.  Along the way, we sketch a solution to the dark matter-baryogenesis coincidence problem with two populations of PBHs, which leads to a double-peaked GW signal. Our results underscore the importance of probing the ultra high frequency GW frontier.}

\begin{document}
\maketitle

\section{Introduction}
\label{sec:introduction}

Gravitational waves (GWs) provide a unique window to the physics of the early Universe. Current detector designs are able to probe GWs  in the nHz-kHz frequency range, leaving the regime of ultra high frequency GWs, in the MHz-GHz range, relatively unexplored. It is precisely at these ultra-high frequencies, however, that one expects very early cosmological events to leave their imprints: very early phase transitions, oscillons, preheating, cosmic strings, etc. This fact has led to a recent surge of interest among particle theorists, cosmologists, and the experimental community in proposing methods to search for GWs at these frequencies. Fully understanding the cosmological and particle physics models driving such proposals is an important aspect of this endeavor. For a comprehensive review of this theoretical and experimental frontier, we refer to \cite{Aggarwal:2020olq, Franciolini:2022htd}. 

The early Universe phenomena listed above all occur below the Big Bang Nucleosynthesis (BBN) bound in the $(h_c, f_{\rm GW})$ plane utilized to depict this frontier (we refer to Fig. 2 of \cite{Aggarwal:2020olq}). Here, $f_{\rm GW}$ denotes the frequency, while $h_c$ denotes the dimensionless characteristic strain relevant to the normalized energy density for stochastic GWs, which we will define carefully in the main text. The extremely low strain relevant to this regime makes it a very challenging target for experimentalists. While primordial black holes (PBHs) have also been studied in this regime, it is in their avatar as stable late-Universe exotic compact objects that they have made their appearance: through inspirals, mergers, or capture (we refer to Fig. 1 of \cite{Aggarwal:2020olq}). We also refer to \cite{Caldwell:2022qsj, Bird:2016dcv, Clesse:2016vqa, Sasaki:2016jop, Kouvaris:2018wnh, Ali-Haimoud:2017rtz, Guo:2019sns, Guo:2017njn} for studies of late-Universe PBHs in the context of lower frequencies attainable at current or near-future detectors.

PBHs that evaporate before BBN could  also be a target for MHz-GHz GW detectors. The source in this case would be the second order GWs produced when primordial scalar fluctuations responsible for PBH formation re-enter the horizon\footnote{This should be contrasted to the GWs produced due to Hawking evaporation itself, which produces GWs at much higher and more challenging frequencies \cite{Anantua:2008am, Dong:2015yjs}.} \cite{1967PThPh..37..831T, Mollerach:2003nq, Ananda:2006af, Baumann:2007zm, Acquaviva:2002ud, Yuan:2021qgz, Domenech:2021ztg, Pi:2020otn, Kohri:2018awv, Espinosa:2018eve, Braglia:2020eai, Inomata:2019ivs, Inomata:2016rbd, Inomata:2018epa}.  In particular, a spell of ultra slow-roll dynamics during the inflationary phase could result in a local enhancement in the spectrum of scalar perturbations, leading to PBH formation \cite{Hertzberg:2017dkh, Ozsoy:2018flq, Franciolini:2022pav, Cicoli:2018asa, Zhang:2021vak, Gao:2021lno}. While scalar, vector, and tensor fluctuations decouple at linear order, at second order the story is different: GWs can be produced with an amplitude that scales quadratically with the scalar perturbation.  Since the amplitude of scalar perturbations is required to be enhanced by orders of magnitude for PBH formation, this significantly enhances the detection prospects of such GWs compared to the primordial signals from inflation. On the other hand, the peak frequency scales with the PBH mass as $f^{\rm peak}_{\rm GW} \sim M^{-1/2}_{\rm PBH}$; for PBH masses between $0.1$ g and $10^4$ g, this falls within the MHz-GHz range\footnote{The upper bound comes from avoiding a spell of PBH domination, as we will see later.}. 

The question then is: do evaporating PBHs connect to problems of interest for particle physicists? Here the answer is very much to the affirmative. The connection of evaporating PBHs to baryogenesis has a long history \cite{Hawking:1975vcx, 1976ZhPmR..24...29Z, 1976ApJ...206....8C, PhysRevD.19.1036, Turner:1979bt, 1980PhLB...94..364G, Alexander:2007gj, Baumann:2007yr, Fujita:2014hha, Hook:2014mla,Hamada:2016jnq, Morrison:2018xla, Bernal:2022pue, Datta:2020bht}. Similarly, dark matter (DM) \cite{Sandick:2021gew, Allahverdi:2017sks, Bell:1998jk, Lennon:2017tqq, Gondolo:2020uqv, Bernal:2021yyb, Cheek:2021odj, Cheek:2021cfe} and dark radiation \cite{Arbey:2021ysg, Hooper:2019gtx, Masina:2021zpu,Cheek:2022dbx} produced by evaporating PBHs have been studied by many groups (we select only a few references from the vast literature). It becomes possible, then, to study the interconnections of ultra-high frequency GW signals and baryogenesis, DM, and dark radiation induced by PBHs.

The purpose of this paper is to explore the interconnections of PBH-induced baryogenesis and GWs between MHz-GHz frequencies\footnote{We leave the interconnections of high frequency GWs with DM and dark radiation from evaporating  light PBHs for future work. The interplay of DM with evaporating PBHs has been explored by several groups -- we mention \cite{Allahverdi:2017sks, Gondolo:2020uqv}. We further note that the interconnection between \textit{stable} PBH DM in the current Universe and low frequency induced GWs has been studied by many authors, for instance recently in~\cite{Kozaczuk:2021wcl, Agashe:2022jgk}. We refer to the review \cite{Domenech:2021ztg, Escriva:2022duf} and references therein. Complementary tests of high-frequency GWs with the GW memory effect in the low frequency range are studied in~\cite{McNeill:2017uvq,Lasky:2021naa}.}. We briefly outline the main steps and our results, beginning with a few preliminary remarks about baryogenesis in general. The matter-anti-matter asymmetry of the Universe is usually explained by the dynamical process of baryogenesis \cite{Dine:2003ax}, which generates the baryon density $\Omega_{B}h^2 = 0.02237 \pm 0.0001$ \cite{Planck:2018nkj}. In the context of PBHs, baryogenesis proceeds as follows: a heavy species $X$ is produced via Hawking evaporation and subsequently decays to Standard Model particles through $CP$ and $B$ violating operators, inducing baryogenesis. 

The final baryon asymmetry is determined by $(i)$ the initial abundance of PBHs; $(ii)$ the number of $X$ particles produced by evaporation per PBH and $(iii)$ the particle physics input in the form of the amount of baryon number violation per decay of $X$, which we call $\gamma_{CP}$. We study the frequency and amplitude of the GWs produced by second order effects as these parameters are varied, requiring that the observed baryon asymmetry is satisfied. There are several criteria that need to be satisfied. Firstly, one must be careful to delineate the regions of parameter space where a matter- (in this case, PBH-) dominated cosmology occurs. A PBH-dominated phase would typically result in GWs peaking in a different frequency range and will be studied in future work. Secondly, one must be careful to avoid washout of the final asymmetry, which can be imposed in this case by the requirement that the mass $m_X$ of $X$ exceed the temperature of the thermal bath at the evaporation time of the PBHs. These features are displayed in Fig.~\ref{fig:betaRD}.

We do not explore the particle physics model-building (in other words, how $X$ couples to the Standard Model) in any great detail, given the plethora of available models in the literature (for completeness, we outline a supersymmetric embedding of the model in the Appendix, based on previous work by one of the authors). Some generic features that we uncover are as follows. Since baryogenesis occurs by the interference of tree-level and one-loop decays of $X$, the asymmetry per decay $\gamma_{CP}$ is typically a simple loop factor $\sim \mathcal{O}(1/16\pi^2)$ for $\mathcal{O}(1)$ Yukawa couplings of $X$. As such, we take a benchmark value of $\gamma_{CP} = 10^{-2}$ for our results. If one lowers the value of $\gamma_{CP}$ to, for example, $\mathcal{O}(10^{-5})$, one needs a higher initial abundance of PBHs to achieve the observed asymmetry. However, one then runs into a PBH-dominated phase, restricting the parameter space where potentially observable GWs can be obtained. This is displayed in the bottom panel of  Fig.~\ref{fig:betaRD}. 

We find that the range of PBH masses relevant for our scenario is $0.1$ g to $\sim 3 \times 10^{4}$ g, with the upper bound coming from the threshold of hitting a PBH-dominated cosmology, assuming $\gamma_{CP} = 10^{-2}$, while the lower bound is from Hubble rate during inflation which is constrained by Cosmic Microwave Background (CMB) observations ~\cite{Planck:2018jri}. 
We calculate the resulting energy density of GWs as a function of the frequency for various PBH masses. These results are shown in Fig.~\ref{fig:GWzoomin}. The strain at the peak frequency of the GW spectrum is plotted against the peak frequency in Fig.~\ref{fig:GWstrain} for different choices of $m_X$. In these figures, we also depict a conceptual design employing the inverse Gertsenshtein effect to probe GWs in this regime\footnote{This was proposed in~\cite{gertsenshtein1962wave}. We provide this reach for the purpose of illustration only; clearly a lot of work needs to be done to probe ultra high-frequency GWs. In additional to the laboratory searches for high-frequency GWs using the inverse Gertsenshtein effect, astrophysical observations of converted electromagnetic signals are also studied in~\cite{Domcke:2020yzq, Kushwaha:2022twx}}. 

To get a sense of the strain values involved, we can provide a summary of our results. The strain at the peak frequency scales as 
\be \label{strainscaling1}
h_c \, \sim \, 5.5 \times 10^{-27} \, \left(\frac{1 \,{\rm MHz}}{f^{\rm peak}_{\rm GW}}\right) \, \log\left[2.0 \times 10^{18} \, \left(\frac{1 \,{\rm MHz}}{f^{\rm peak}_{\rm GW}}\right)^{2} \, \left(\frac{\gamma_{CP}}{10^{-2}}\right)^{2} \, \left(\frac{8.7 \times 10^{-11}}{Y_B}\right)^{2} \right]^{-1},
\ee
for 
\be \label{mxcomp1}
5.41 \times 10^2 \, \left(\frac{f^{\rm peak}_{\rm GW}}{1\,{\rm MHz}}\right)^3\,\,{\rm GeV} \lesssim m_X \lesssim 1.33 \times 10^8 \, \left(\frac{f^{\rm peak}_{\rm GW}}{1\,{\rm MHz}}\right)^2 \,\,{\rm GeV},
\ee
and
\be \label{strainscaling2}
h_c \, \sim \, 5.5 \times 10^{-27} \, \left(\frac{1 \,{\rm MHz}}{f^{\rm peak}_{\rm GW}}\right) \, \log\left[ 5.9 \times 10^{14} \, \left(\frac{f^{\rm peak}_{\rm GW}}{1 \,{\rm MHz}}\right)^{6} \, \left(\frac{10^{9} \, {\rm GeV}}{m_X}\right)^{4} \, \left(\frac{\gamma_{CP}}{10^{-2}}\right)^{2} \, \left(\frac{8.7\times10^{-11}}{Y_B}\right)^2 \right]^{-1},
\ee
for 
\be \label{mxcomp2}
1.33 \times 10^8 \, \left(\frac{f^{\rm peak}_{\rm GW}}{1\,{\rm MHz}}\right)^2 \,\,{\rm GeV} \lesssim m_X \lesssim 1.04 \times 10^8 \, \left(\frac{\gamma_{CP}}{10^{-2}}\right)^{1/2} \, \left(\frac{8.7\times10^{-11}}{Y_B}\right)^{1/2} \, \left(\frac{f^{\rm peak}_{\rm GW}}{1 \,{\rm MHz}}\right)^{5/2} \,\,{\rm GeV}.
\ee
Here,  $Y_B=n_B/s \simeq 8.7\times10^{-11}$. The lower  bound on $m_X$ in Eq.~(\ref{mxcomp1}) comes from avoiding washout. Eqs.~(\ref{strainscaling1}) and (\ref{strainscaling2}) are of course valid as long as one stays within a radiation-dominated cosmology. The frequency of GWs is determined by the PBH mass that solves baryogenesis in a radiation-dominated universe. We found the lowest frequency is about $1.64\times(10^{-2}/\gamma_{CP})\times[Y_B/(8.7\times10^{-11})]~{\rm MHz}$, which indicates MHz-GHz GWs in the parameter space of baryogenesis models.

Finally, we explore a scenario where the DM-baryogenesis coincidence problem is addressed by two populations of PBHs: one responsible for baryogenesis, and the other stable and comprising DM today. The coincidence is then primarily driven by the  ratio of the masses and abundances of the two populations. The striking feature of this scenario is a double-peaked GW spectrum.

The rest of our paper is organized as follows. In Sec.~\ref{sec:PBHmassfunction}, we discuss PBH formation from scalar perturbations. In Sec.~\ref{sec:Baryogenesis}, we discuss the mechanism of generating baryon asymmetry with PBH Hawking radiation in the early universe. GW signals generated by curvature perturbations are discussed in Sec.~\ref{sec:GravitationalWaves}. In additional to baryogenesis, we study the possibility of solving the cosmological coincidence problem with a bimodal PBH mass function in Sec.~\ref{sec:CoincidenceProblem}. We conclude the main result of this study in Sec.~\ref{sec:conclusion}.

\section{PBHs mass function}
\label{sec:PBHmassfunction}

In this Section, we introduce our notation and discuss the  PBH mass function that will be used to address baryogenesis. The PBH mass function at time $t$ is defined as the ratio between the PBH energy density per logarithmic mass interval and the total radiation energy density at the formation time, $t_i(M)$,
\bea
\frac{d\beta(M_{\rm PBH},t_i,t)}{d\log M_{\rm PBH}}=
\left\{
        \begin{array}{ll}
        \frac{a(t)}{a(t_i)}\frac{1}{\rho_\text{rad}(t_i)}\frac{d\rho(M_{\rm PBH})}{d\log M_{\rm PBH}}  & \qquad {\rm (Radiation-dominated),}\\
        \\
        \frac{1}{\rho'_\text{rad}(t,t_i)}\frac{a^3(t_i)}{a^3(t)}\frac{d\rho(M_{\rm PBH})}{d\log M_{\rm PBH}} & \qquad {\rm (PBH-dominated).}
        \end{array}
\right.
\eea
Here $d\rho(M_{\rm PBH})/d\log M_{\rm PBH}=M_{\rm PBH}\times dn(M_{\rm PBH})/d\log M_{\rm PBH}$ and $dn(M_{\rm PBH})/d\log M_{\rm PBH}$ is the number density of PBH per logarithmic mass interval at the formation time. $a(t)$ is the scale factor. If the universe is always radiation-dominated, the radiation energy density $\rho_\text{rad}$ scales as $a^{-4}$ while the PBH energy density scales as $a^{-3}$. If there exists a PBH-dominated epoch, the energy density of radiation is determined by the reheating temperature from PBH evaporation. We denote the radiation energy density after a PBH-dominated epoch as $\rho'_{\rm rad}$. The properties of $\rho'_{\rm rad}$ is discussed in detail in Sec.\ref{sec:PBHMD}.

The fraction of energy density collapsed into PBHs of mass $M_{\rm PBH}$ can be calculated from the fraction of Hubble patches that undergoes gravitational collapse in the existence of overdensities at the horizon size $R$. We assume the overdensities are generated by curvature perturbations $P_{\zeta}(k)$ from the inflationary epoch. At the re-entry time of a $k$-mode, $k=aH$, overdensities are generated at the scale $R=k^{-1}$ and the amplitude of the density contrast $\delta=\delta\rho/\rho$ is assumed to follow a Gaussian distribution
\bea
p(\delta)=\frac{1}{\sqrt{2\pi}\sigma_0} \, e^{-\frac{\delta^2}{2\sigma^2_0}}.
\eea
The mean of the $\delta$ distribution is zero and the variance $\sigma_0$ can be calculated with 
\bea
\sigma_0^2(k=R^{-1})=\displaystyle{\int_0^{\infty}}\frac{{\rm d}k'}{k'}\frac{16}{81}(k'R)^4W^2(k',R)P_\zeta(k'),
\label{eq:sigma0}
\eea
where a Gaussian window function is used to smooth the power spectrum contribution,  
\bea
W(k,R)=\exp\left[-\frac{(kR)^2}{2}\right].
\eea
Note that large variance is also generated at neighbouring scales of an enhanced curvature perturbation mode as a result of the integral in Eq.(\ref{eq:sigma0}).

Gravitational collapse into a PBH happens in patches where the overdensity $\delta$ is larger than the threshold $\delta_c$. We take $\delta_c=1/3$ when the energy density of the Universe is dominated by radiation. The number density of PBHs can be calculated by evaluating the probability of a Hubble patch having $\delta>\delta_c$ at the horizon re-entry time. The fraction of patches that collapse into PBHs is
\bea
\frac{n_{\rm PBH}}{n_{\rm patch}}&=&\int_{\delta_c}^{\infty}d\delta \, \frac{1}{\sqrt{2\pi}\sigma_0}e^{-\frac{\delta^2}{2 \sigma_0^2}}\\ \nonumber
&=&\frac{1}{2} \, {\rm Erfc}\left(\frac{\delta_c}{\sqrt{2}\sigma_0}\right).
\eea

 We further assume that a fraction $\gamma$ of the horizon mass will become the final PBH mass, $M_{\rm PBH}=\gamma M_{R=k^{-1}}$. The horizon mass can be expressed as:
\bea
M_{R}=2\times10^{4}~{\rm g} \, \left(\frac{k}{10^{21}~{\rm Mpc}^{-1}}\right)^{-2}.
\label{eq:MR}
\eea
Then the mass function at the formation time $t=t_i$ can be written as: 
\bea
\frac{d\beta(M,t_i,t_i)}{d\log M}&=&\frac{M_{\rm PBH} \, n_{\rm PBH}}{M_{R} \, n_{\rm patch}}\nonumber\\
&=&\frac{\gamma}{2}{\rm Erfc}\left(\frac{\delta_c}{\sqrt{2}\sigma_0}\right).
\label{eq:dbetadlogM}
\eea

If we take the curvature perturbation to be monochromatic, the power spectrum can be parameterized with two parameters, the amplitude $A_\zeta$ and the peak location $k_0$,
\bea
P_{\zeta}(A_\zeta,k_0)=A_\zeta \, \delta\left(\log \left(\frac{k}{k_0}\right)\right), 
\label{eq:deltaPzeta}
\eea
while the variance is simplified to  
\bea
\sigma^2_0(k)=A_\zeta \, \frac{16}{81} \, 
\left(\frac{k_0}{k}\right)^4 \, \exp\left[-\left(\frac{k_0}{k}\right)^2\right].
\label{eq:Sigma0sqDeltaPS}
\eea

A monochromatic power spectrum would generate a peaked PBH mass distribution, whose initial abundance can be defined in the similar way:
\begin{equation}
\beta(M_{\rm PBH})=\frac{\rho(M_{\rm PBH})}{\rho_{\rm rad}(t_i(M_{\rm PBH}))}= \frac{n_{\rm PBH} \times M_{\rm PBH}}{\rho_{\rm rad}(t_i(M_{\rm PBH}))}.
\label{eq:massfunctiondelta}
\end{equation}
In following sections, we assume a single PBH population of fixed mass is responsible for baryogenesis, and use Eq.(\ref{eq:massfunctiondelta}) as an approximation of Eq.(\ref{eq:dbetadlogM}) in the numerical analysis.

\section{Baryogenesis by Hawking Evaporation}
\label{sec:Baryogenesis}
A population of PBHs which undergoes  Hawking evaporation can emit any particle in the spectrum which is lighter than their instantaneous temperature. In this section, we summarize baryogenesis through the $CP$ and baryon number violating decay of a beyond-Standard Model particle $X$ produced by Hawking evaporation of PBHs\footnote{This idea was initially introduced as a viable production mechanism of heavy bosons predicted by Grand Unified Theories ~\cite{Turner:1979zj, Turner:1979bt}.}. 

Our focus is on the production of the PBHs and the interplay of the  concomitant GW signal with the mass $m_X$ and the baryon-number-to-entropy density $Y_B$. For explicit particle physics models of the subsequent baryon asymmetry obtained by the decay of $X$, we refer to \cite{Baumann:2007yr, Smyth:2021lkn, Morrison:2018xla} and   Appendix~\ref{sec:susy}, where we give details of some supersymmetric embeddings based on \cite{Allahverdi:2013tca, Allahverdi:2010rh, Dutta:2010sg, Allahverdi:2010im}.
If PBHs dominate the energy density of the Universe prior to their evaporation, they can initiate an early matter- (PBH-) dominated era. We will study both the case where PBHs evaporate in a radiation-dominated Universe, as well as when they themselves come to dominate the energy density. 

\subsection{Evaporation of PBHs in a Radiation-Dominated Universe}

Density fluctuations, represented as $\delta$, grow after entering the cosmological horizon in the early Universe. If that re-entering happens in a radiation-dominated epoch, the over-density may overcome the pressure and collapse into a PBH. The collapse can happen if $\delta>\omega$ where $\omega=p/\rho$ is the equation of state parameter~\cite{Carr:1975qj}. If the energy density of the Universe is dominated by radiation, $\omega=1/3$. In a radiation-dominated era where the Universe is filled  with $g_\star(T)$ relativistic degrees of freedom sharing the same temperature $T$, and the energy density is given by 
\begin{equation}
   \rho_\text{rad}(T)=\frac{\pi^2}{30}g_\star(T)T^4,
\end{equation}
and the Hubble expansion rate follows from Friedmann equation as:
\begin{equation}
    H^2(T)=\frac{8\pi}{3M_\text{Pl}^2}\rho_\text{rad}(T).
\end{equation}
If PBHs form within this era at some formation time denoted by $t_i$, then the mass of PBHs is proportional to the horizon mass at that time, i.e.,
\begin{equation}
   M_\text{PBH}=\frac{4\pi}{3}\gamma\rho_\text{rad}(T_i)H^{-3}(T_i),
\end{equation}
where $\gamma\sim\omega^{3/2}\approx 0.2$. The temperature of the Universe at formation time, therefore, can be expressed as:
\begin{equation}
   T_i(M_\text{PBH})=\frac{\sqrt{3}\,5^{1/4}}{2\pi^{3/4}}\sqrt{\gamma}\frac{1}{g^{1/4}_\star(T_i)}\frac{M_\text{Pl}^{3/2}}{\sqrt{M_\text{PBH}}}.
\end{equation}
PBHs which form during radiation domination gain negligible spin~\cite{DeLuca:2019buf}. So in this study we assume that all the PBHs are Schwarzschild (non-rotating).  PBHs lose their mass by Hawking evaporation~\cite{Hawking:1975vcx} during which they emit particles that are lighter than their instantaneous horizon temperature given by:
\begin{equation}
T_\text{PBH}(t)=\frac{M_\text{Pl}^2}{8\pi M_\text{PBH}(t)}.
\label{eq:temp}
\end{equation}
The lifetime of a PBH, $\tau_\text{PBH}$, with initial mass $M_{\rm{PBH}}$ can be shown~\cite{Gondolo:2020uqv} to be equal to:
\begin{equation}
\tau_\text{PBH}=\frac{10240\pi}{g_\star(T_\text{PBH})}\frac{M^3_\text{PBH}}{M_\text{Pl}^4}.
\label{eq:lifetimePBH}
\end{equation}
Since in a radiation-dominated Universe, $H(t)=1/(2t)$, by combining Friedmann equation with the lifetime of PBHs, one can obtain the temperature of the Universe at their evaporation time as:
\begin{equation}
T_{\rm eva}\,=\,\frac{\sqrt{3}\,g^{1/4}_\star(T_\text{PBH})}{64\sqrt{2}\,5^{1/4}\pi^{5/4}}\frac{M_\text{Pl}^{5/2}}{M_\text{PBH}^{3/2}}.
\label{eq:Trevap}
\end{equation}
PBHs may give rise to an early matter-dominated era. Due to the fact that PBHs and radiation are diluted differently with the expansion of the Universe, the ratio of energy density of PBHs to the energy density of radiation increases with time, i.e.,  $\rho_\text{PBH}(t)/\rho_\text{rad}(t)\propto a$. 
Therefore, a transition from a radiation-dominated era to a matter-dominated one is inevitable if PBHs have long enough lifetimes.
The critical initial abundance of PBHs, $\beta_c$, to start an early matter-dominated era can be evaluated by requiring that PBHs evaporate after the early equality time, $t_\text{early-eq}$, at which $\rho_\text{PBH}(t_\text{early-eq})/\rho_\text{rad}(t_\text{early-eq})\sim 1$. 
The temperature of the Universe at early equality time, $T_\text{early-eq}$, can be obtained from:
\begin{equation}
\frac{\rho_\text{PBH}(T_\text{early-eq})}{\rho_\text{rad}(T_\text{early-eq})}=\frac{\rho_\text{PBH}(T_i)}{\rho_\text{rad}(T_i)}\frac{T_i}{T_\text{early-eq}}=\beta_c\frac{T_i}{T_\text{early-eq}}\sim 1.
\end{equation}
An early matter-dominated epoch is unavoidable provided that $t_\text{early-eq}\lesssim t_\text{eva}$. This is equivalent to $\beta\geq\beta_c$, where 
\begin{equation}
\beta_c=\frac{T_\text{eva}}{T_i}=\sqrt{\frac{g_\star(T_\text{PBH})}{10240\pi\gamma}}\frac{M_\text{Pl}}{M_\text{PBH}}\simeq2.8\times 10^{-6}\left(\frac{g_\star(T_\text{PBH})}{106.8}\right)^{1/2}\left(\frac{0.2}{\gamma}\right)^{1/2}\left(\frac{1 \text{g}}{M_\text{PBH}}\right).
\label{eq:MD}
\end{equation}

If PBHs never dominate the energy density of the Universe prior to their Hawking evaporation, i.e., $\beta<\beta_c$, then they will evaporate in a radiation-dominated era at some time denoted by $t=t_\text{eva}$. Among other particles, PBHs emit the  beyond-Standard Model particle $X$ with a lifetime of $\tau_X$ which is added to the spectrum to explain baryogenesis by its decay at time  $t=t_\text{eva}+\tau_X$. It is customary to define the baryon asymmetry $\gamma_{CP}$ produced per decay of $X$, given by 
\begin{equation}
   \gamma_{CP}=\sum_i B_i\frac{\Gamma(X\rightarrow f_i)-\Gamma(\bar{X}\rightarrow \bar{f_i})}{\Gamma_X},
\end{equation}
where $B_i$ is the baryon number of the particular final state $f_i$, and $\Gamma_X$ is the total decay width of $X$. $f_i$ and $\bar{f_i}$ denote a  particle and its anti-particle, respectively. Baryogenesis is obtained by the interference between tree-level and one-loop decays of $X$. The exact value of $\gamma_{CP}$ depends on the particle physics model; however, for $\mathcal{O}(1)$ Yukawa couplings of $X$ to other particles, one obtains
\be
\gamma_{CP} \, \sim \, \frac{1}{16\pi^2},
\ee
coming from a loop suppression. As a benchmark, we use the value $\gamma_{CP} = 10^{-2}$ in our work, and keep the scaling with $\gamma_{CP}$ explicit throughout.

Hawking evaporation of a population of PBHs with mass $M_{\rm PBH}$ and initial abundance $\beta$, where $\beta<\beta_c$, gives rise to baryon-number-to-entropy density, $Y_B$, equal to
\begin{eqnarray}
    Y_B&=&\frac{n_B(t_0)}{s(t_0)}=\gamma_{CP}\nonumber\frac{n_X(t_\text{eva}+\tau_X)}{s(t_\text{eva}+\tau_X)}=\gamma_{CP}\frac{n_X(t_\text{eva})}{s(t_\text{eva})}=\gamma_{CP} N_X\frac{n_\text{PBH}(t_{i})}{s(t_{i})}\\
    &=&\gamma_{CP}\beta N_X\frac{1}{M_{\rm PBH}}\frac{\rho_\text{rad}(t_{i})}{s(t_{i})}=\frac{3}{4}\frac{g_\star(T_i)}{g_{\star,S}(T_i)}\gamma_{CP}\beta N_X \frac{T_{i}(M_{\rm PBH})}{M_{\rm PBH}},
    \label{eq:BRD}
\end{eqnarray}
where $g_{\star,S}(T_i)$ counts the number of relativistic degrees of freedom contributing to the entropy of the Universe, and $N_X$, is the total number of particles $X$ emitted by one PBH given by~\cite{Gondolo:2020uqv}
\begin{equation}
N_X= \frac{15\zeta(3)}{\pi^3}\frac{g_X}{g_\star(T_\text{PBH})}\left\{
        \begin{array}{ll}
            \frac{8M_\text{PBH}^2}{M_\text{Pl}^2}& \quad m_X<T_\text{PBH}, \\
            \\
           \frac{1}{8\pi^2}\frac{M_\text{Pl}^2}{m_X^2} & \quad m_X>T_\text{PBH}\,,
        \end{array}
    \right.
    \label{eq:NX}
\end{equation}
if $X$ is a boson. Here, $g_X$ denotes the number  of degrees  of freedom of $X$. We set $g_X = 1$ in all our figures, although we keep the scaling with $g_X$ explicit in our calculations.  The total number of fermionic species is $N_F=\frac{3}{4}\frac{g_F}{g_B}N_B$.

\begin{figure}[t]
  \centering
    \includegraphics[width=0.4\textwidth]{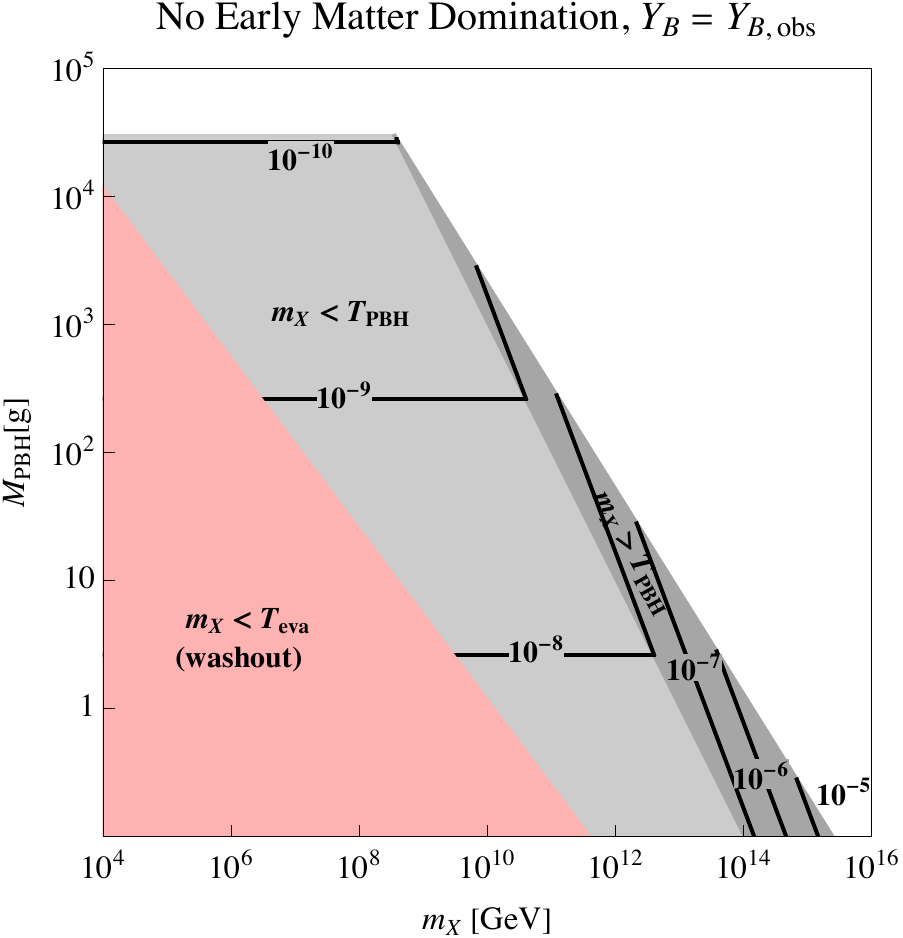}\hspace{3mm}
  \includegraphics[width=0.40\textwidth]{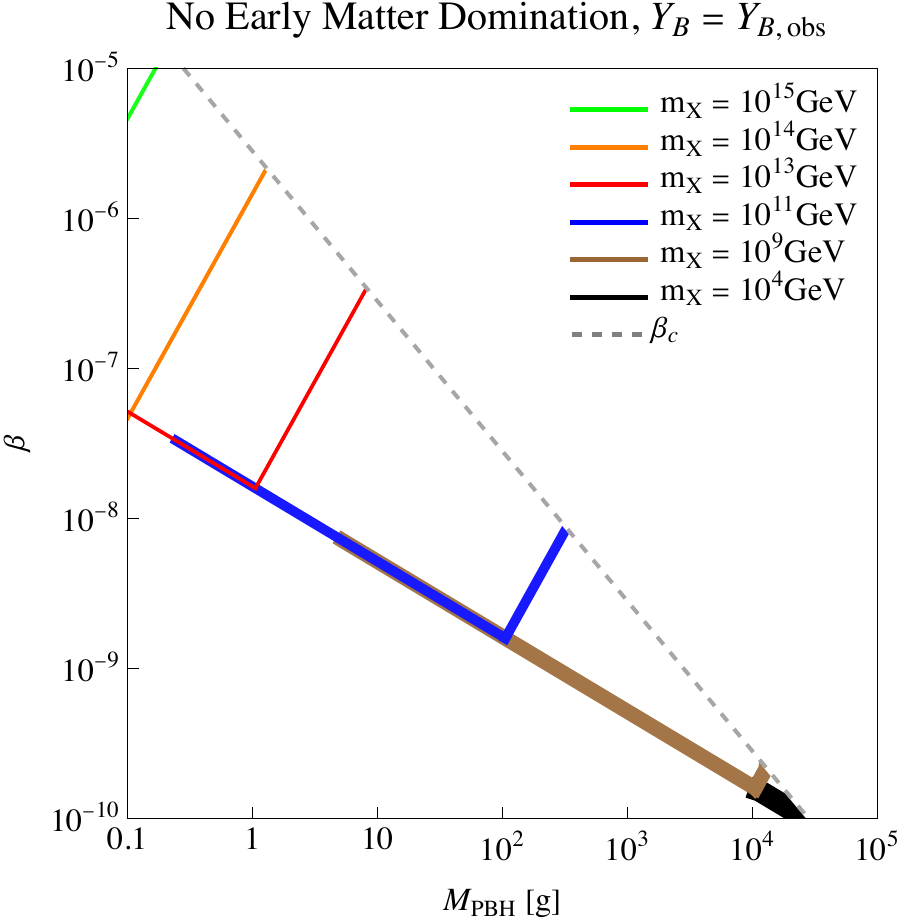}\\
   \includegraphics[width=0.40\textwidth]{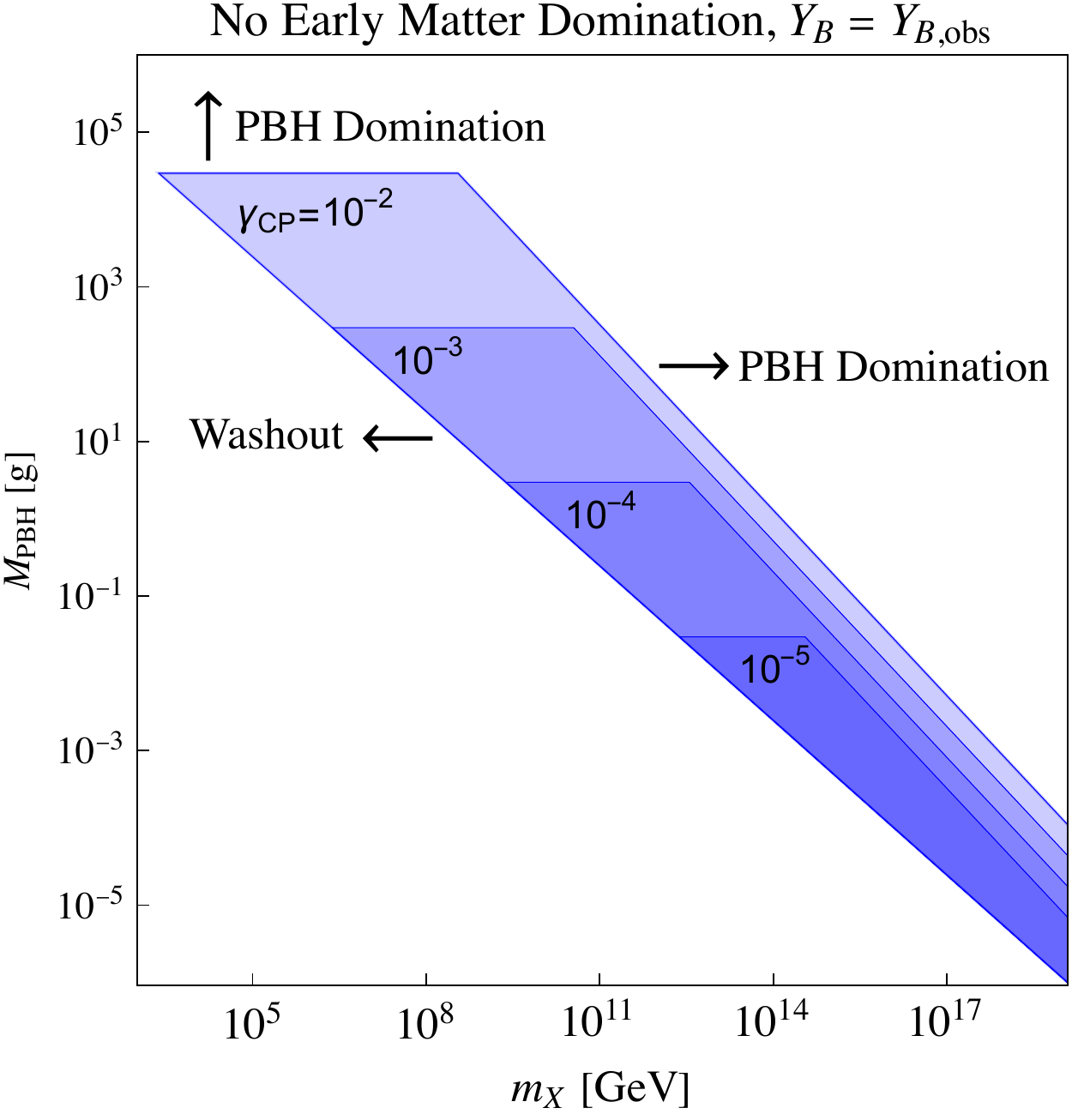}\hspace{3mm}
  \caption{\textit{Top-left panel}: Contours of $\beta$ are drawn on the $(m_X, M_{\rm PBH})$ plane requiring the observed value of $Y_B$ in the radiation-dominated case $(\beta <\beta_c)$. We use the value $\gamma_{CP} = 10^{-2}$. The black lines show the  initial abundance of PBHs that can explain the baryon asymmetry of the Universe without giving rise to an early matter-dominated era $(\beta <\beta_c)$. The light gray area corresponds to $m_X<T_\text{PBH}$ while the dark gray area marks the region in which $m_X>T_\text{PBH}$.  The white region corresponds to $\beta>\beta_c$ and a PBH-dominated era. The   pink region is ruled out due to the requirement $m_X > T_{\rm eva}$, needed to avoid washout.  \textit{Top-right panel}: Contours of $m_X$ are drawn on the $(\beta, M_{\rm PBH})$ plane requiring the observed value of $Y_B$ in the radiation-dominated case $(\beta <\beta_c)$. We use the value $\gamma_{CP} = 10^{-2}$. The different colors 
   show the values of different choices of $m_X$. The gray dashed line marks $\beta_c$.  \textit{Bottom panel}: The 
   parameter space on the plane of $M_{\rm PBH}$ vs. $m_X$ that can produce the correct $Y_B$ in a radiation-dominated cosmology, for different value of $\gamma_{CP}$. Each lighter shade of blue is a superset that also includes the deeper colors with smaller values of $\gamma_{CP}$. The dependence on $\gamma_{CP}$ is discussed in the main text. Regions that are ruled out due to washout or that feature a PBH-dominated cosmology are indicated. We display PBH masses  $M_{\rm PBH} \lesssim 10^{-5}$ g for illustrative purposes: to indicate the regime where especially small values of $\gamma_{CP}$ are operational.}
  \label{fig:betaRD}
\end{figure}

\begin{figure}[t]
  \centering
    \includegraphics[width=0.4\textwidth]{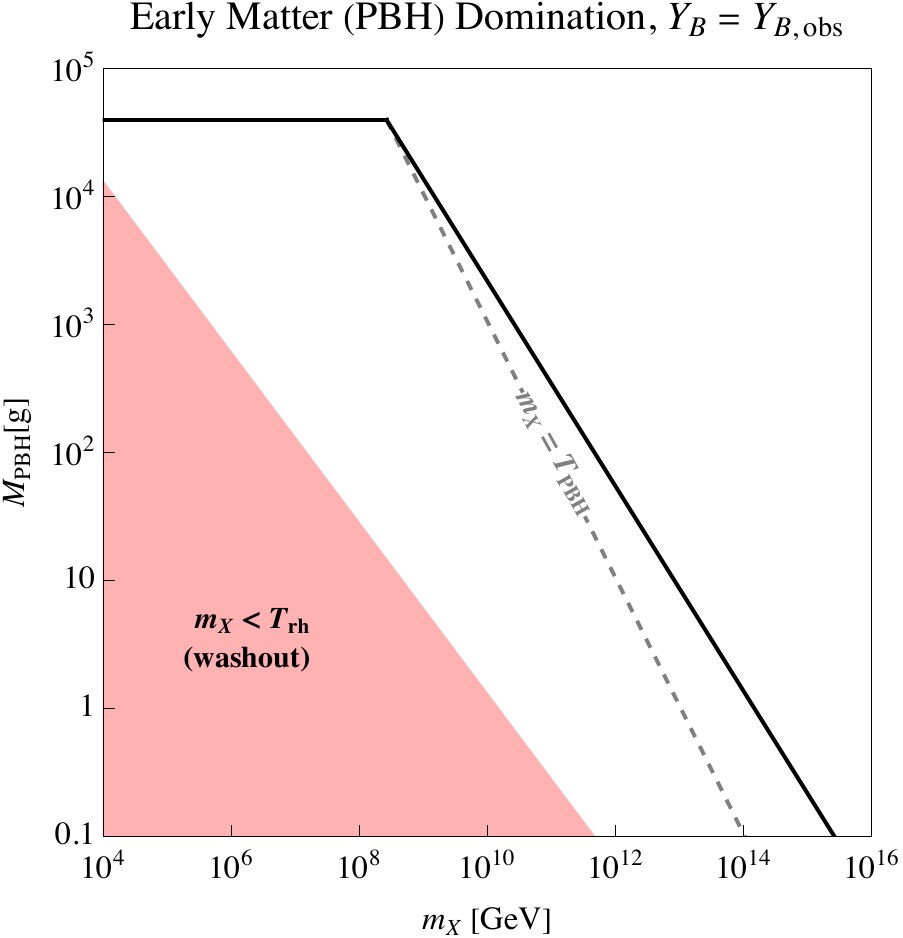}
  \caption{Black line shows $M_\text{PBH}$, as a function of $m_X$, that explains observed value of $Y_B$ (for $\gamma_{CP}=10^{-2}$) when PBHs come to dominate the energy density of the Universe before their evaporation ($\beta\geq\beta_c$). Along the gray dashed line we have $m_X=T_\text{PBH}$. The pink region is excluded to evade washout by requiring $m_X > T_\text{rh}$.}
  \label{fig:betaMD}
\end{figure}

The initial abundance of PBHs which leads to a value $Y_B$ of the baryon-number-to-entropy density can be obtained as 
\begin{equation}
\beta= \frac{\pi^{15/4}}{45\sqrt{3}\,5^{1/4}\zeta(3)}\frac{g_\star(T_\text{PBH})g_\star^{1/4}(T_{i})}{g_X}\gamma^{-1/2}\gamma_{CP}^{-1}Y_B\left\{
        \begin{array}{ll}
            \left(\frac{M_\text{Pl}}{M_{\rm PBH}}\right)^{1/2}& \quad m_X<T_\text{PBH}, \\
            \\
            64\pi^2\left(\frac{M^3_{\rm PBH} m_X^4}{M_\text{Pl}^7}\right)^{1/2}& \quad m_X>T_\text{PBH}\,,
        \end{array}
    \right.
    \label{eq:betallimit}
\end{equation}

Avoiding an early matter-dominated phase, i.e., $\beta<\beta_c$ is equivalent to an upper bound on the mass of PBHs, $M_{\rm PBH}< M_{c}$ where 
\begin{equation} \label{masscrit}
M_{c}=\left\{
        \begin{array}{ll}
            \frac{1215\sqrt{5}\zeta^2(3)}{2048\pi^{17/2}}\frac{g_X^2}{g_\star(T_\text{PBH})g_\star^{1/2}(T_i)}\gamma_{CP}^2Y^{-2}_B M_\text{Pl}& \quad m_X<T_\text{PBH}~\left(M_{\rm PBH}<\frac{M_\text{Pl}^2}{8\pi m_X}\right), \\
            \\
           \frac{3\times 5^{3/10}\zeta^{2/5}(3)}{16\times 2^{3/5}\pi^{5/2}}\frac{g_X^{2/5}}{g_\star^{1/5}(T_\text{PBH})g_\star^{1/10}(T_i)}\gamma_{CP}^{2/5}Y^{-2/5}_B\left( \frac{M_\text{Pl}^9}{m_X^4}\right)^{1/5} & \quad m_X>T_\text{PBH}~\left(M_{\rm PBH}>\frac{M_\text{Pl}^2}{8\pi m_X}\right)\,,
        \end{array}
    \right.
\end{equation}

These conditions can be combined as
\begin{equation}
\left\{
        \begin{array}{ll}
            M_{\rm PBH}<\text{min}\left\{\frac{1215\sqrt{5}\zeta^2(3)}{2048\pi^{17/2}}\frac{g_X^2}{g_\star(T_\text{PBH})g_\star^{1/2}(T_i)}\gamma_{CP}^2Y^{-2}_B M_\text{Pl},\, \frac{M_\text{Pl}^2}{8\pi m_X}\right\}& \quad m_X<T_\text{PBH}, \\
            \\
        \frac{M_\text{Pl}^2}{8\pi m_X} < M_{\rm PBH}< \frac{3\times 5^{3/10}\zeta^{2/5}(3)}{16\times 2^{3/5}\pi^{5/2}}\frac{g_X^{2/5}}{g_\star^{1/5}(T_\text{PBH})g_\star^{1/10}(T_i)}\gamma_{CP}^{2/5}Y^{-2/5}_B\left( \frac{M_\text{Pl}^9}{m_X^4}\right)^{1/5} & \quad m_X>T_\text{PBH}\,.
        \end{array}
    \right.
\end{equation}

Following~\cite{Zyla:2020zbs}, the observational value that we use for $Y_B$ is 
\begin{equation}
    Y_B=\frac{n_B}{s}\simeq\frac{2.515\times 10^{-7}\,\text{cm}^{-3}}{2891.2\,\text{cm}^{-3}}\simeq 8.7\times10^{-11}.
\end{equation}

It is clear from Eq.~(\ref{masscrit}) that the critical PBH mass to avoid matter domination is determined mainly by the observed value of $Y_B$ and the value of $\gamma_{CP}$. While this is completely true in the case that $m_X < T_\text{PBH}$, for the case that $m_X > T_\text{PBH}$ this statement is true up to $\sim \left(M_\text{Pl}/m_X\right)^{1/5}$. The top two panels of Fig.~\ref{fig:betaRD} show the parameter space of PBHs that can explain baryon asymmetry of the Universe without transitioning to an early matter-dominated era. In the top-left panel of Fig.~\ref{fig:betaRD}, the black lines depict the initial abundance of PBHs, i.e., $\beta$, in the $(m_X, M_{\rm PBH})$ plane. PBHs with the presented initial abundance can emit enough $X$ particle to explain baryon asymmetry of the Universe. In the light gray region, the $X$ particle is lighter than the initial temperature of PBHs, while in the dark gray area $X$ is heavier than the temperature of the PBHs. The border between these two regions marks the line $T_\text{PBH}=m_X$. Within the white region, $\beta$ exceeds the critical value $\beta_c$ and leads to a PBH-dominated Universe. 

To avoid washout of the baryon asymmetry generated by the decay of $X$, we require
\be
m_X \, > \, T_{\rm eva}\,\,,
\ee
where the temperature of the Universe at evaporation time, $T_{\rm eva}$, is given by Eq.~(\ref{eq:Trevap}).
This constraint is shown in the left panel of Fig.~\ref{fig:betaRD} by the  pink region. While avoiding an early PBH-dominated era leads to an upper bound on PBH mass, i.e., $M_\text{PBH}\lesssim 3\times 10^4\,\text{g}$, evading washout gives rise to a lower limit on PBH mass which varies with $m_X$ (only for $m_X<T_\text{PBH}$).

We note that the minimum evaporation temperature to avoid a PBH-dominated phase corresponds to $M_\text{PBH}\simeq 3\times 10^{4}\,\text{g}$ and is  given by $2.4\,\text{TeV}$. This also implies that $m_X$ has to be greater than this value. We therefore have
\be
T_{\rm baryo} \, \geq \, 2.4 \,\,{\rm TeV}\,\,.
\ee
To avoid subsequent sphaleron washout, the decays of $X$ should be $B-L$ violating.
 
The top right panel of Fig.~\ref{fig:betaRD} displays limits on the mass of $X$ particle in the $(M_{\rm PBH}, \beta)$ plane. The different colors correspond to different values of $m_X$. The dotted line shows $\beta_c$. For any given color, the parameter space along the diagonal line corresponds to the case where $m_X < T_\text{PBH}$, while the sharp uptick corresponds to the case where $m_X > T_\text{PBH}$. Along the diagonal, a given color corresponding to a fixed value of $m_X$ starts at a point corresponding to the condition $m_X > T_{\rm eva}$. Along the uptick, the color stops at the point where $\beta = \beta_c$, i.e., where matter domination takes over.

The bottom panel of Fig.~\ref{fig:betaRD} shows the parameter space of $\gamma_{CP}$ on the $(m_X, M_{\rm PBH})$ plane in a radiation-dominated cosmology that gives the correct baryon asymmetry. For a given shade of blue, the horizontal cut-off as well as the cut-off on the right diagonal come from avoiding PBH domination. The cut-off on the left diagonal comes from avoiding washout. These features are displayed in the figure. We also note that we go down to PBH masses of $\lesssim 10^{-5}$ g to indicate the regime where especially small values of $\gamma_{CP}$ are operational. 

\subsection{Evaporation of PBHs in a Matter- (PBH-) Dominated Universe}
\label{sec:PBHMD}

If the initial abundance of PBHs is large enough, i.e., $\beta\geq\beta_c$, they can dominate the energy density of the Universe and subsequently induce an early matter- (PBH-) dominated epoch\footnote{For a general review of early matter-dominated eras, we refer to \cite{Allahverdi:2020bys, Kane:2015jia}.}. In other words, the evaporation of PBHs reheats the Universe for the second time and initiates a secondary radiation-dominated epoch.
The quick equilibration of the SM particles emitted by PBHs sets the secondary reheating temperature of the Universe.

By applying the Friedemann equation very close to the Hawking evaporation of PBHs, their energy density can be expressed in terms of their lifetime as
\begin{equation}
H^2(\tau)=\left(\frac{2}{3\tau_{\rm PBH}}\right)^2=\frac{8\pi\rho(\tau_{\rm PBH})}{3M_\text{Pl}^2},
\label{eq:friedmannMD}
\end{equation}
where the lifetime of black hole, $\tau_\text{PBH}$, is given by Eq.~(\ref{eq:lifetimePBH}).
The reheating temperature of the Universe, $T_\text{rh}(\tau_{\rm PBH})$, given instantaneous equilibration, equals to:
\begin{equation}
T_\text{rh}(\tau_{\rm PBH})=\frac{1}{32\sqrt{2}\times 5^{1/4}\pi^{5/4}}g_{\star}^{1/4}(T_\text{rh})\left(\frac{M_\text{Pl}^5}{M^{3}_{\rm PBH}}\right)^{1/2}.
\label{eq:TVtauMD}
\end{equation}
Following Eq.~(\ref{eq:BRD}), baryon-number-to-entropy density, is equal to:
\begin{equation}
    Y_B=\gamma_{CP} N_X\frac{n_\text{PBH}(\tau_{\rm PBH})}{s(\tau_{\rm PBH})}=\gamma_{CP} N_X\frac{1}{M_{\rm PBH}}\frac{\rho_\text{PBH}(\tau_{\rm PBH})}{s(\tau_{\rm PBH})}=\frac{3}{4}\frac{g_\star(T_\text{rh})}{g_{\star,S}(T_\text{rh})}\gamma_{CP} N_X \frac{T_\text{rh}(\tau_{\rm PBH})}{M_{\rm PBH}},
    \label{eq:BMD}
\end{equation}
where $N_X$ is given by Eq.~(\ref{eq:NX}). Eq.~(\ref{eq:BMD}) shows that when PBHs dominate the energy density of the Universe prior to their evaporation, the abundance of emitted particles is independent of initial abundance of PBHs~\cite{Baumann:2007yr}.

One can solve Eq.~(\ref{eq:BMD}) for  $M_{\rm PBH}$ as a function of $Y_B$ and $m_X$. When $m_X<T_\text{PBH}$, the result is independent of $m_X$ and there is only one mass for PBHs, given by:
\begin{equation}
    M_{\rm PBH}=\frac{405\sqrt{5}\zeta(3)^2}{512\pi^{17/2}}\frac{g_X^2}{g_\star^{3/2}(T_\text{rh})}\gamma_{CP}^2Y_B^{-2}M_\text{Pl},
\end{equation}
while for $m_X>T_\text{PBH}$, we get:
\begin{equation}
    M_\text{PBH}=\frac{3^{4/5}\times 5^{3/10}\zeta^{2/5}(3)}{16\times 2^{1/5}\pi^{5/2}}\frac{g_X^{2/5}}{g_\star^{3/10}(T_\text{rh})}\gamma_{CP}^{2/5}Y_B^{-2/5}\left(\frac{M_\text{Pl}^9}{m_X^4}\right)^{1/5}.
\end{equation}

Fig.~\ref{fig:betaMD} displays the mass of PBHs as a function of $m_X$ when they dominate the Universe before their evaporation ($\beta\geq\beta_c$) and the $X$ particles explain baryon asymmetry of the Universe for $\gamma_{CP}=10^{-2}$.
For $10^4\,\text{GeV}\lesssim m_X\lesssim 2.7\times 10^8\,\text{GeV}$ where $m_X<T_\text{PBH}$, we obtain $M_\text{PBH}\simeq 4\times 10^4\,\text{g}$.
By increasing the $m_X$ and entering $m_X>T_\text{PBH}$ region, $M_{\rm PBH}$ needs to be decreased to explain baryon asymmetry.
 
Our focus in this work will be on GWs produced in a radiation-dominated Universe. The interplay of baryogenesis and GWs produced  during a PBH-dominated phase will be explored in the future. Nevertheless, we point out some general features of this correlation in subsection \ref{sec:GWfromMD}.

\section{MHz - GHz Gravitational Waves and Interplay with Baryogenesis}

\label{sec:GravitationalWaves}

In this section, we provide details of the calculation of the GWs coming from PBH formation. We will calculate the frequency and amplitude of the signal that is compatible with baryogenesis.

The curvature perturbations responsible for PBH formation generate GW signals~(see~\cite{Kohri:2018awv, Inomata:2018epa} for the analytical formulae of induced GWs). We note that GW production is independent of whether PBHs are actually formed. The GWs are mostly produced at the moment of horizon crossing and diluted with the expansion of the universe, becoming a   stochastic GW background today. We focus on the production of GWs from scalar perturbations that enter the horizon during a radiation-dominated era. GWs are generated at the second order of the scalar mode and the abundance of them can be estimated as $\Omega_{\rm GW}\simeq A^2_{\zeta}$ where $A_\zeta\sim10^{-2}$ is the amplitude of the scalar perturbations generating PBHs. The peak frequency of induced GWs is fixed by the horizon re-entry time since sub-horizon source modes decay quickly when the universe is radiation-dominated.

The energy density of GWs normalized to the critical density per logarithmic $k$ interval at a conformal time $\eta$  is given by
\begin{equation}
\Omega_{\rm GW}(\eta,k)=\frac{1}{24}\left(\frac{k}{a(\eta)H(\eta)}\right)^2P_h(\eta,k)\,.\label{eq.GW}
\end{equation}
The GW power spectrum can be written as 
\bea
P_h(\eta,k)&\simeq&2\int^{\infty}_{0}dt\int^{1}_{-1}ds\left(\frac{t(t+2)(s^2-1)}{(t+s+1)(t-s+1)}\right)^2\nonumber\\
&&\times I^2(s,t,k\eta)P_\zeta(u k)P_\zeta(v k),
\label{eq:GWPh}
\eea
where the dimensionless variables are $u=\frac{t+s+1}{2}$ and $v=\frac{t-s+1}{2}$. In a radiation-dominated Universe, perturbations decay quickly after horizon re-entry. The GWs are mostly produced at the re-entry time and evolve to constant values in the sub-horizon limit. Then the $I^2$ term can be written as 
\bea
I^2(s,t,k\eta)&=&\frac{288(s^2+t(t+2)-5)^2}{k^2\eta^2(t+s+1)^6(t-s+1)^6}\bigg(\frac{\pi^2}{4}\Big(s^2+t(t+2)-5\Big)^2\Theta(t-(\sqrt{3}-1))\nonumber\\
&&\qquad\qquad+\Big(-(t+s+1)(t-s+1)+\frac{1}{2}(s^2+t(t+2)-5)\log\Big|\frac{t(t+2)-2}{3-s^2}\Big|\Big)^2\bigg),\nonumber\\
\eea
where $\Theta(.)$ is the Heaviside function.

The GW density today $\Omega_{\rm GW,0}$ can be calculated by red-shifting GWs together with other radiations from a time $\eta_s$ when the GWs stop growing and become a fixed fraction of the total radiation energy density $\Omega_{\rm rad}$,
\bea
\Omega_{\rm GW, 0}(k)
&=&\Omega_{\rm rad,0} \, \frac{\Omega_{\rm GW}(\eta_{\rm eq},k)}{\Omega_{{\rm rad, eq}}}\nonumber\\
&=&2 \, \Omega_{\rm rad,0} \, \frac{H^2(\eta_s) \, a^4(\eta_s)}{H^2_{\rm eq} \, a^4_{\rm eq}} \, \Omega_{\rm GW}(\eta_s,k) \nonumber\\
&=&2 \, \Omega_{\rm rad,0} \, \frac{a^4(\eta_s) \, \rho_{\rm rad}(\eta_s)}{a^4_{\rm eq} \, 2 \, \rho_{\rm rad,eq}} \, \Omega_{\rm GW}(\eta_s,k)\nonumber\\
&=&\Omega_{\rm rad,0} \, \frac{a^4(\eta_s) \, g_{\star}(\eta_s) \, T^4(\eta_s)}{a^4_{\rm eq} \, g_{\star,{\rm eq}} \, T^4_{\rm eq}} \, \Omega_{\rm GW}(\eta_s,k)\nonumber\\
&=&\left( \frac{g_{\star}(\eta_s)}{g_{\star,{\rm eq}}} \right) \, \left( \frac{g_{\star, S}(\eta_s)}{g_{\star, S,{\rm eq}}} \right)^{-\frac{4}{3}} \, \Omega_{\rm rad,0} \, \Omega_{\rm GW}(\eta_s,k) \nonumber\\
&=&0.83 \, \left(\frac{g_{\star}(\eta_s)}{10.75}\right)^{-\frac{1}{3}} \, \Omega_{\rm rad,0} \, \Omega_{\rm GW}(\eta_s,k).
\label{eq:GWdeltaToday}
\eea
In the second line, we used $\Omega_{{\rm rad, eq}}=1/2$, $H^2=8\pi G\rho_{\rm rad}/3$ and $\rho_{\rm GW}\propto a^4$. In the fifth line, we used conservation of entropy $g_{\star,S,{\rm eq}} a^3_{\rm eq} T^3_{\rm eq}=g_{\star, S}(\eta_s) a^3(\eta_s) T^3(\eta_s)$. In the last line, we used $g_{\star, S,{\rm eq}}=3.91$, $g_{\star,{\rm eq}}=3.36$ and $g_{\star}(\eta_s)=g_{\star, S}(\eta_s)$.
The energy density of radiation today is $\Omega_{\rm rad,0}=8.5\times10^{-5}$. The effective massless degrees of freedom $g_{\star}(\eta_s)$ at $\eta_s$ is $106.75$ for the PBH mass of our interest.

When the perturbation is monochromatic as defined in Eq.($\ref{eq:deltaPzeta}$), the GW density at production time can be written as 
\bea
\Omega_{\rm GW}(k=rk_0)&=&\frac{3}{64} A^2_\zeta \, r^2 \, \left(\frac{4-r^2}{4}\right)^2 \, (2-3r^2)^2 \nonumber\\
&\quad\times&\left[\left(4+(3r^2-2)\log \left|\frac{3r^2-4}{3r^2}\right|\right)^2+\pi^2(3r^2-2)^2\Theta\left(\frac{2}{\sqrt{3}}-r\right)\right] \Theta(2-r),\nonumber\\
\label{eq:GWdelta}
\eea
which is proportional to $A^2_\zeta$. The GW frequency is related to $k$ as:
\be
\frac{f_{\rm GW}}{{\rm Hz}}=1.546\times10^{-15} \frac{k}{{\rm Mpc}^{-1}}\,\,.
\label{eq:GWktof}
\ee
The $\Omega_\text{GW}$ spectrum in Eq.(\ref{eq:GWdelta}) is scale-invariant, and its peak location is at $r\to2/\sqrt{3}$ where the  $\Omega_{\rm GW}$ divergence is cut off by $\Theta(\frac{2}{\sqrt{3}}-r)$.  The GW density is  proportional to the square of the power spectrum amplitude from the nature that GWs are induced at the second order in scalar perturbations.
We can then write down the relation between the PBH mass and the frequency of the GW spectrum peak that comes from the perturbation as
\bea
f_{\rm GW}^{\rm peak}\simeq 2.82\times \left(\frac{M_{\rm PBH}}{10^{4}~{\rm g}}\right)^{-\frac{1}{2}}{\rm MHz}.
\label{eq:fGWpeak}
\eea
The GW signal from the formation of  PBHs peaks at $f_{\rm GW}^{\rm peak}\sim ~ \mathcal{O}({\rm MHz})$, which is much higher than the observation window of future space-based GW proposals. 

We can now calculate the GW spectrum that is consistent with baryogenesis. The procedure is as follows. From Eq.~(\ref{eq:betallimit}), we obtain the value of $\beta$ required to give the correct value of $Y_B$, for a given $M_{\rm PBH}$. We fix $\gamma_{CP} = 10^{-2}$ and require $m_X$ to be smaller than $T_{\rm PBH}$ such that $\beta$ is independent of $m_X$ in the light gray region in the top left panel of Fig. \ref{fig:betaRD}. Once $\beta$ is obtained, the power spectrum in Eq.~(\ref{eq:deltaPzeta}) that gives the correct $\beta$ is calculated from the formalism outlined in Section~\ref{sec:PBHmassfunction}. Thereafter, $\Omega_\text{GW}$ is obtained from Eq.~(\ref{eq:GWdeltaToday}).

\begin{figure}[h]
\centering
\includegraphics[scale=0.55]{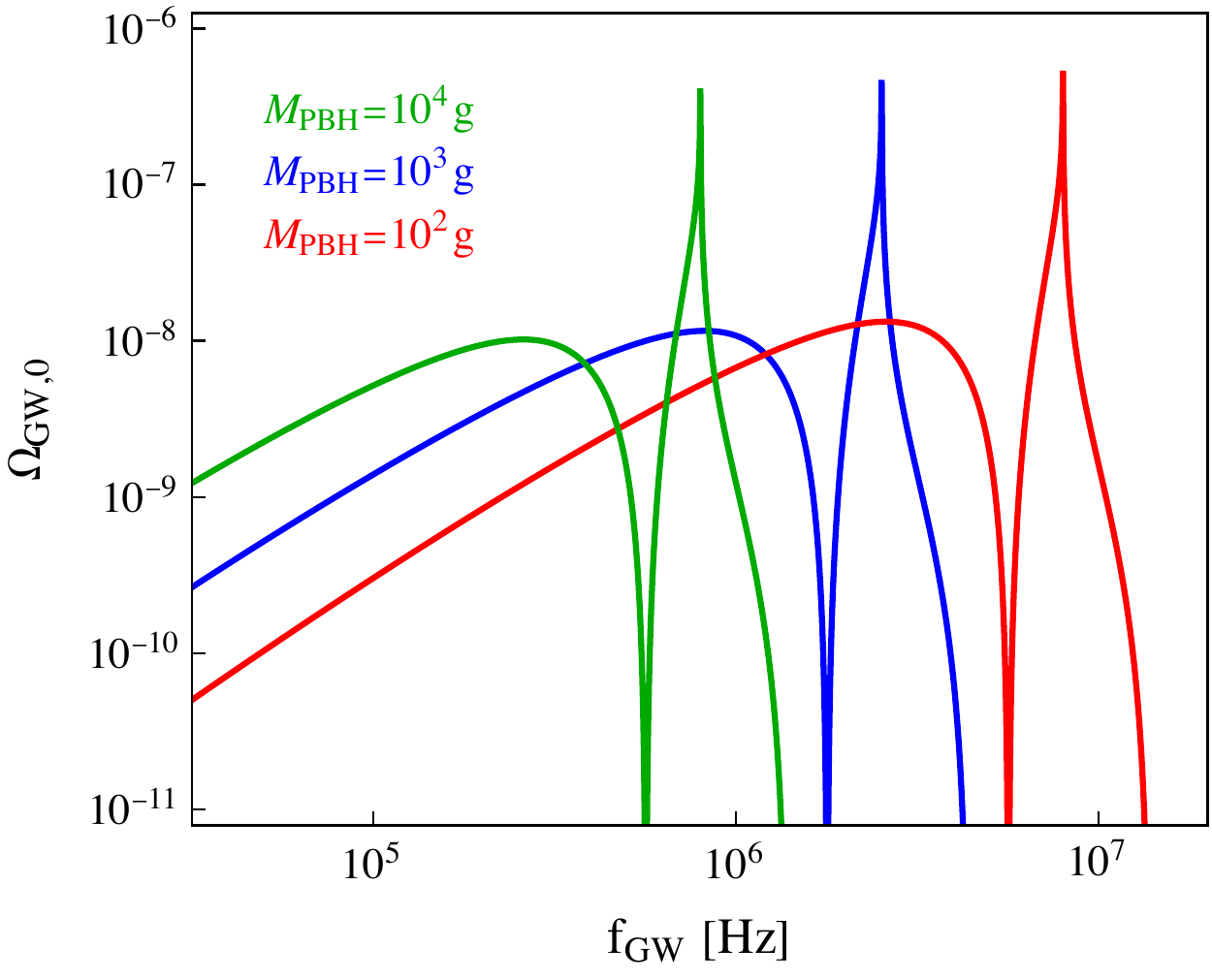}\qquad
\includegraphics[scale=0.55]{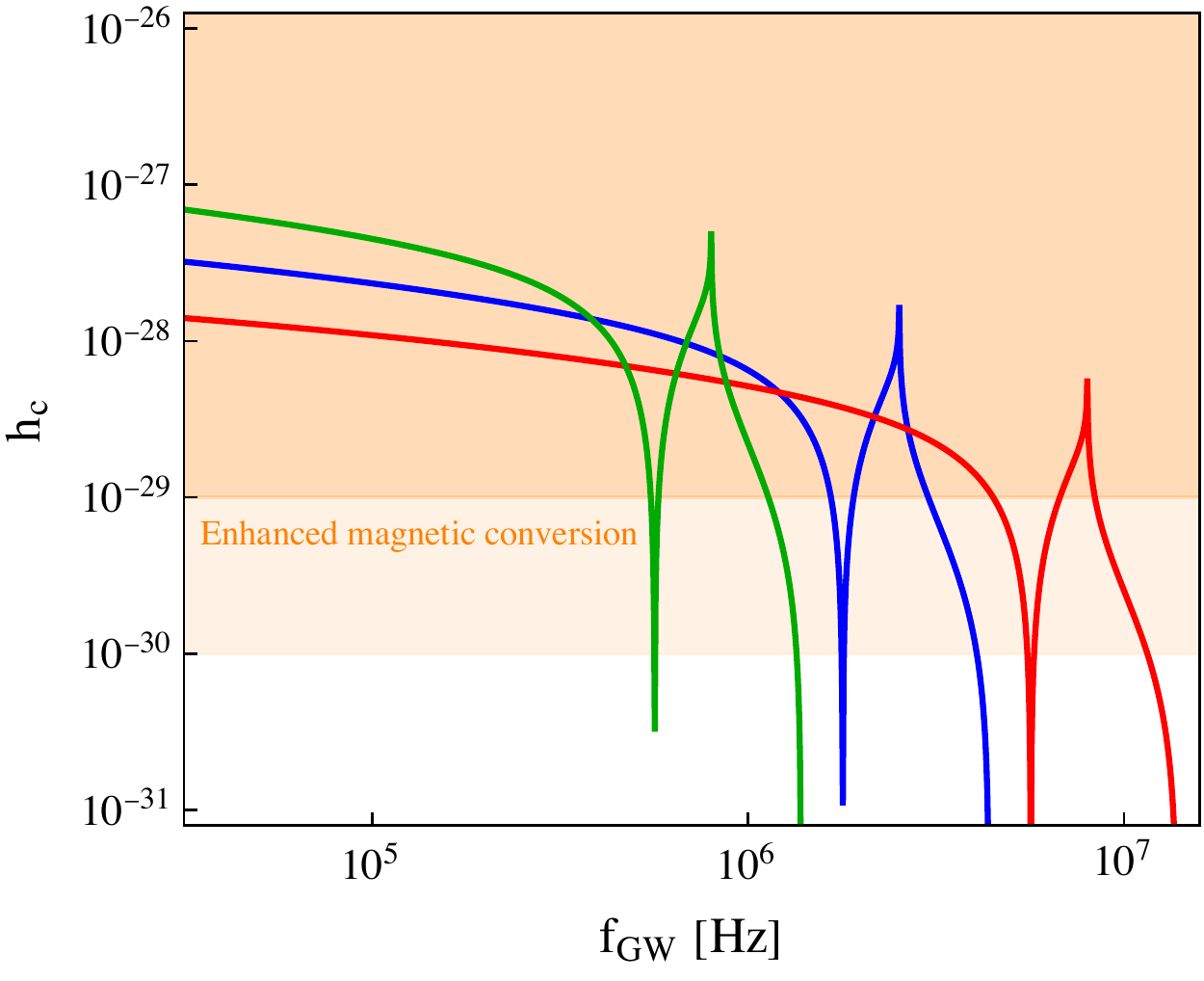}
\caption{{\it Left panel:} The energy density of GWs as a function of the frequency, for three different values of $M_{\rm PBH}$: $10^{2}$ g (red),  $10^{3}$ g (blue), and $10^{4}$ g (green). The observed value of $Y_B$ is obtained for each curve,  with $\gamma_{CP} = 10^{-2}$ and $m_X < T_{\rm BH}$. The details of the calculation are in the text. {\it Right panel:} The strain $h_c$ is shown as a function of the frequency for the three values of $M_{\rm PBH}$ from the left panel. The orange  shaded regions show the  future sensitivity limit of the high frequency GW probe  using the inverse Gertsenshtein effect proposed in ~\cite{gertsenshtein1962wave}. The dark orange shade is a conservative projection of the sensitivity, while the light orange shade is an optimistic projection.}
\label{fig:GWzoomin}
\end{figure}

In Fig.~\ref{fig:GWzoomin} (left panel), we show results for $\Omega_\text{GW}$ for several selections of PBH masses:  $M_{\rm PBH} = 10^2$g, $10^3$g, and $10^4$g. In Fig.~\ref{fig:GWzoomin} (right panel), we display the strain $h_c$ given by
\bea
h_c(f)=\frac{1}{f}\sqrt{\frac{3 \, H^2_0 \, \Omega_{\rm GW}}{4 \, \pi^2}},
\eea
corresponding to the three PBH benchmark masses. Since the strain of induced GWs is proportional to the amplitude of the power spectrum, $h_c\propto A_{\zeta}$, the characteristic strain value is $\sim 10^{-28}$ at $f_{\rm GW}\sim ~ \mathcal{O}({\rm MHz})$. 

\begin{figure}[h]
\centering
\includegraphics[width=0.45\textwidth]{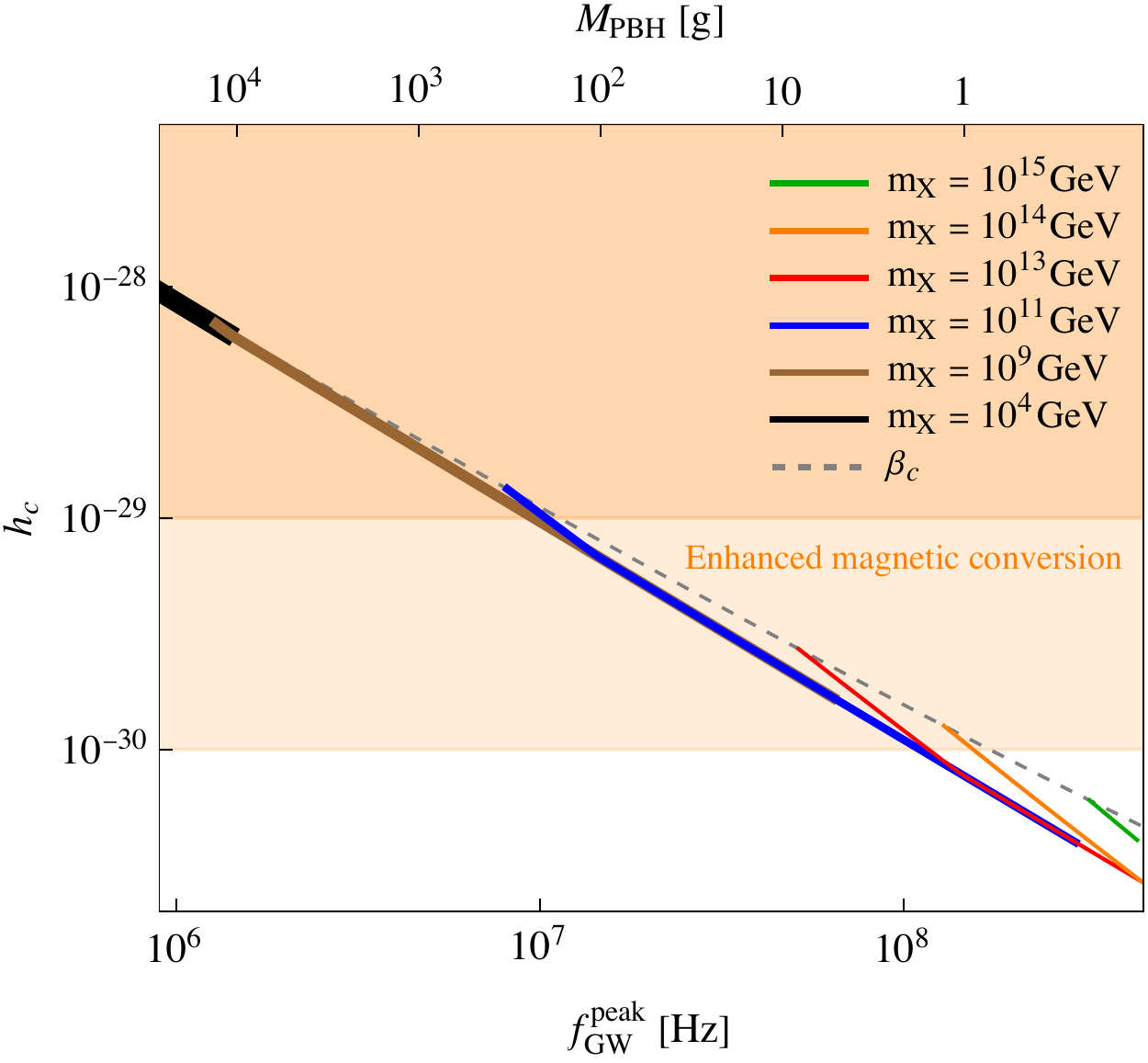}
\caption{The strain at the peak frequency for high frequency GWs coming from light PBHs, as a function of the peak frequency~(lower x-axis) and as a function of the light PBH mass~(upper x-axis). The orange shaded regions show the  future sensitivity limit of the high frequency GW probe  using the inverse Gertsenshtein effect proposed in ~\cite{gertsenshtein1962wave}. The darker orange region is a conservative projection of sensitivity, while the lighter orange region is an optimistic projection.}
\label{fig:GWstrain}
\end{figure}

High frequency GW detection in the MHz - GHz range is a major target of interest for theorists. The prospects and challenges for future detection are discussed in the recent review \cite{Aggarwal:2020olq} (we also refer to \cite{Berlin:2021txa, Domcke:2022rgu} and references therein). One idea that has been proposed in the frequency range of our interest employs the inverse Gertsenshtein effect~\cite{gertsenshtein1962wave}, in which a graviton can be converted resonantly into a photon in the presence of a static background magnetic field. Small strain values are still very difficult to measure because the rate of the inverse Gertsenshtein process is suppressed by $h^2_c$. Instead of measuring the converted photon directly, one can also measure the GW conversion in a region where an auxiliary electromagnetic field is generated by placing a laser beam together with the magnetic field~\cite{Li:2009zzy}. The event rate of the modified measurement is only linear in $h_c$ hence the sensitivity of the proposal can be significantly improved. The initial proposal in~\cite{Li:2009zzy} is aimed at $5~{\rm GHz}$ frequency and a strain strength as small as $h_c \sim10^{-30}$~(See Table 1 of~\cite{Aggarwal:2020olq}). The status of technology development for enhanced magnetic conversion is recently reviewed in~\cite{Ringwald:2020ist} and the sensitivity to $h_c$ is estimated to be $h_c\sim 10^{-29}$ when current state-of-art benchmarks are included. Further improvement is expected with the future development of gyrotrons, single-photon detectors and magnets. In Fig.~\ref{fig:GWzoomin} and Fig.~\ref{fig:GWstrain}, we show a conservative sensitivity projection~($h_c=10^{-29}$) and an optimistic sensitivity projection~($h_c=10^{-30}$) in dark and light orange shade respectively.
We note that the above reach is shown for completeness; clearly, a lot of work needs to be done by the community to probe GWs in this interesting high  frequency range.

\begin{figure}[h]
\centering
\includegraphics[width=0.45\textwidth]{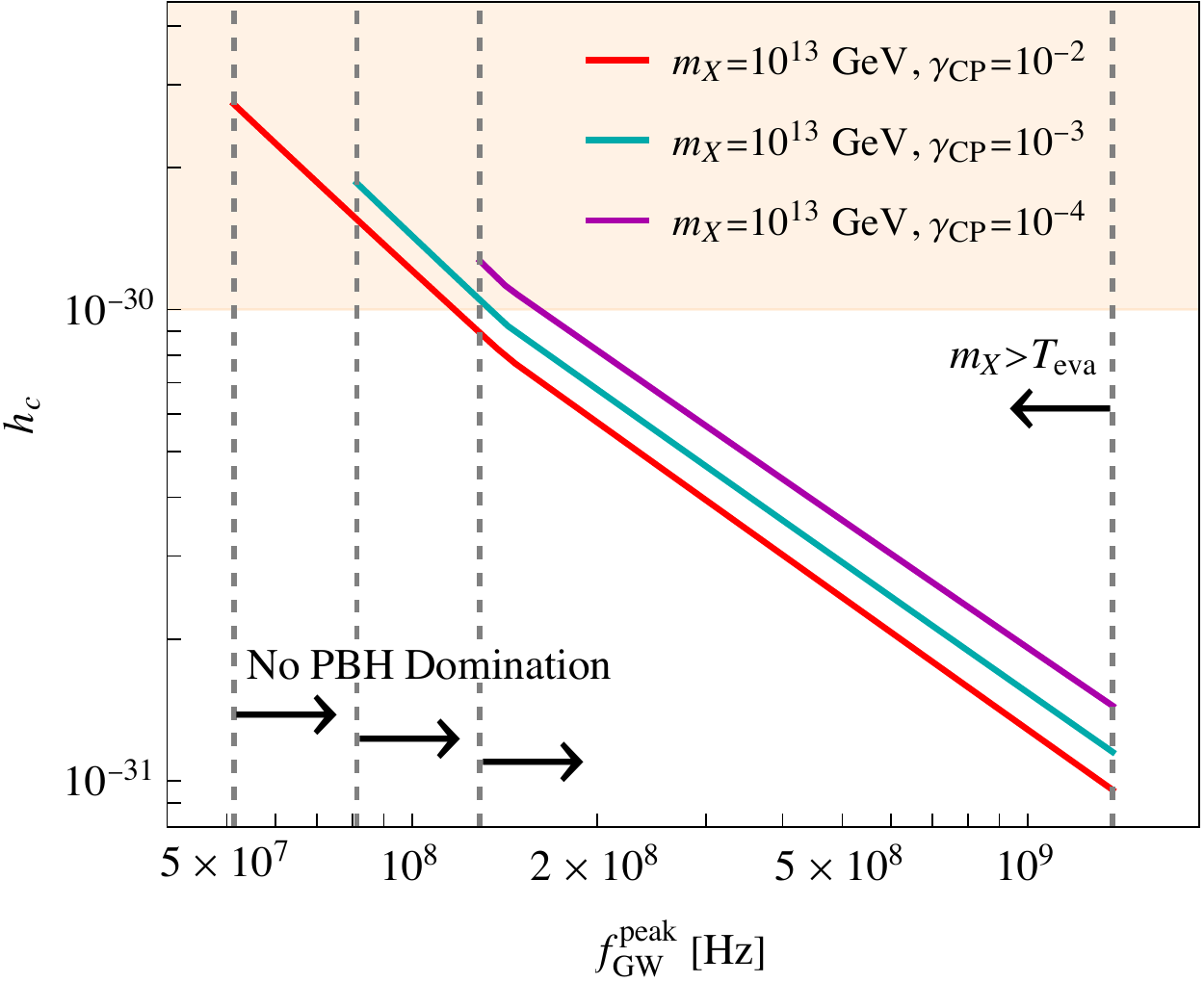}\quad
\includegraphics[width=0.45\textwidth]{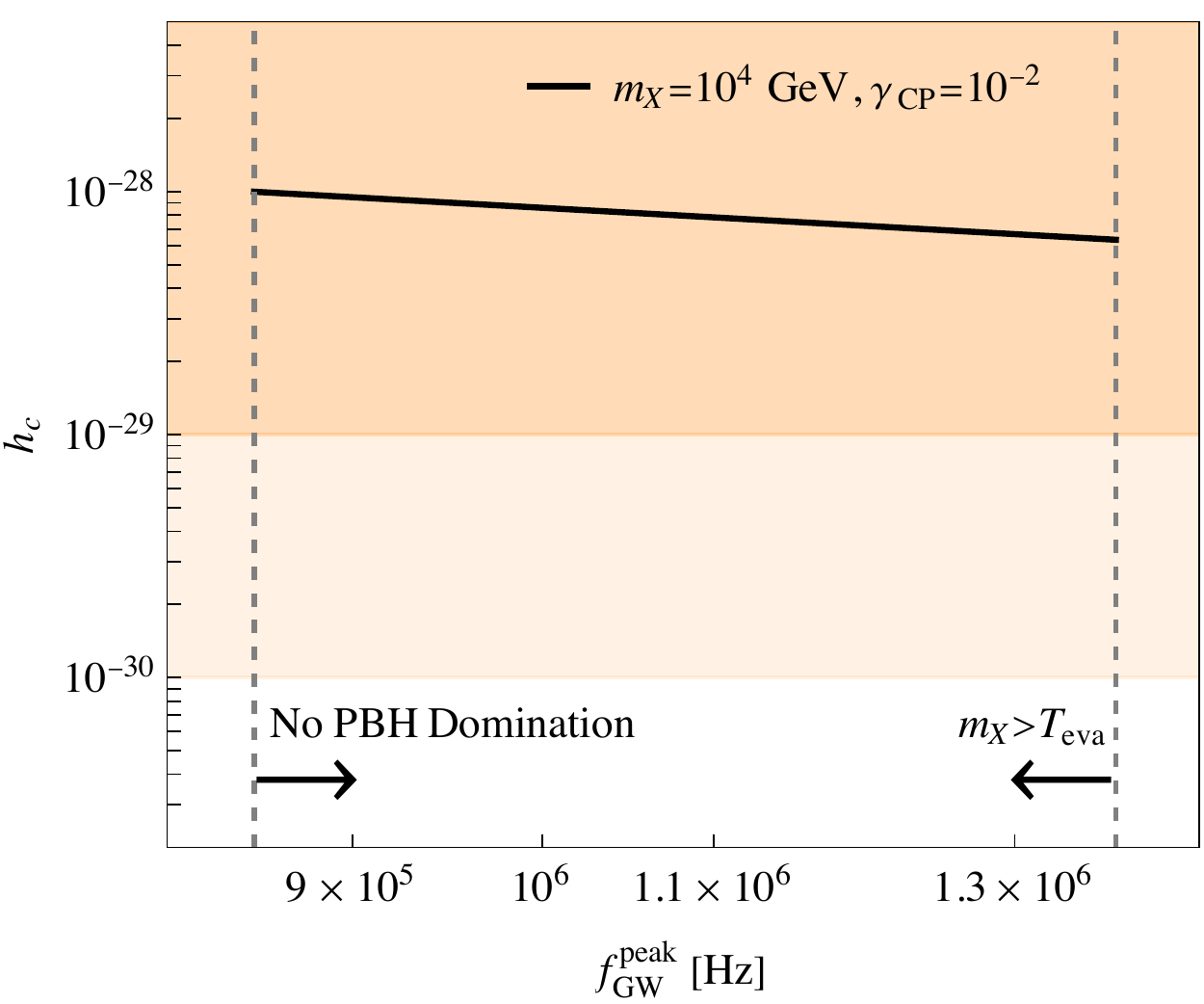}
\caption{{\it Left panel:} The strain at the peak frequency for benchmark models of $m_X=10^{13}~{\rm GeV}$ and different values of $\gamma_{CP}=10^{-2}$ (red), $10^{-3}$ (cyan) and $10^{-4}$ (purple). The vertical dashed lines show the $f^{\rm peak}_{\rm GW}$ regions where successful baryogenesis can take place in a radiation-dominated universe. {\it Right panel:} The strain at the peak frequency for a benchmark model of $m_X=10^{4}~{\rm GeV}$ and $\gamma_{CP}=10^{-2}$. Note that a large $\gamma_{CP}$ value is required for baryogenesis when $m_X$ is small, as is shown in the bottom panel of  Fig.~\ref{fig:betaRD}. }
\label{fig:GWstrainBenchmark}
\end{figure}

In Fig.~\ref{fig:GWstrain}, we study the strain at the peak of the GW spectrum as a function of the peak frequency (lower x-axis) and the  PBH mass (upper x-axis). The different colors correspond to various values of $m_X$, with  the color  scheme following top right panel of Fig.~\ref{fig:betaRD}.  This calculation merits a caveat. As can be seen from Fig.~\ref{fig:GWzoomin}, the GW signal is divergent at the peak, and the peak value for the energy density or strain actually depend on how close one approaches the peak frequency. The basic cause of this is the assumption of a delta function PBH mass function in Eq.~(\ref{eq:massfunctiondelta}), or, equivalently, a delta function power spectrum in Eq.~(\ref{eq:deltaPzeta}). To plot the peak strain in Fig.~\ref{fig:GWstrain}, we instead choose a smoother power spectrum, whose details are provided in Appendix \ref{smoothpsepc}. The relation between $\beta$ and $M_{\rm PBH}$ that gives the correct $Y_B$ is first taken from the top right panel of Fig.~\ref{fig:betaRD}. Then, the amplitude of a delta function power spectrum that gives the correct $\beta$ at a given $M_{\rm PBH}$ is obtained. The delta function is then ``smoothed" with the spectrum that scales like $k^4$, depicted in the top left panel of Fig.~\ref{fig:k4}, keeping the same amplitude at the peak as the delta function. Finally, the GW spectrum is  calculated with the smoothed $k^4$ spectrum and the  value of the strain $h_c$ at the peak frequency is obtained. The process is repeated for different  PBH masses to obtain Fig.~\ref{fig:GWstrain}.

\subsection{Scaling of strain with peak frequency}

We are finally in a position to present our results on the $( f^{\rm peak}_{\rm GW}, h_c )$ plane for different choices of the particle physics parameters $m_X$ and $\gamma_{CP}$. The results are shown in Fig.~\ref{fig:GWstrainBenchmark}. In the left panel, we show the results for a choice of $m_X = 10^{13}$ GeV, for different values of $\gamma_{CP} = 10^{-2}, 10^{-3},$ and $10^{-4}$. The vertical lines impose the conditions to avoid PBH domination and washout. For each value of $\gamma_{CP}$, the curves have two scalings: the steeper one corresponds to $m_X > T_{BH}$, while the less steep portion corresponds to $m_X < T_{BH}$. In the right panel, we display the corresponding result for $m_X = 10^4$ GeV and $\gamma_{CP} = 10^{-2}$. Here, only the regime $m_X < T_{BH}$ applies.

 Semi-analytic behaviors of the strain versus peak frequency curves can be found in the following way.  The PBH abundance $\beta$ needed to generate baryon asymmetry $Y_B$ is calculated with Eq.~(\ref{eq:betallimit}) for chosen $m_X, \gamma_{CP}$, and $M_{\rm PBH}$. The power spectrum amplitude $A_\zeta$ is then obtained from $\beta$ with Eq.~(\ref{eq:dbetadlogM}) and Eq.~(\ref{eq:Sigma0sqDeltaPS}) assuming contribution to baryogenesis is mainly from the peak mass.\footnote{We used the approximation ${\rm Erfc}^{-1}(z)|_{z\to 0} \simeq \sqrt{\frac{1}{2}\log[\frac{2}{\pi z^2}]-\frac{1}{2}\log[\log[\frac{2}{\pi z^2}]]}\sim \sqrt{\frac{1}{2}\log[\frac{2}{\pi z^2}]}$, with which the error is at about the $5\% \,(9\%)$ level for $\beta=10^{-10}\,(10^{-5})$.} The power spectrum $k$-mode can be solved from Eq.~(\ref{eq:MR}) and is related to the GW peak frequency $f^{\rm peak}_{\rm GW}$ in Eq.~(\ref{eq:GWktof}). To estimate the strain and energy density of GWs from the power spectrum, we used the approximation $\Omega_{\rm GW}(\eta_{s},k)\simeq P^2_{\zeta}(k)\simeq A^2_{\zeta}$, which is shown to reproduce the correct order of magnitude at the peak frequency for log-normal power spectra in~\cite{Inomata:2018epa}. The results are:

\be \label{strainscaling1a}
h_c \, \sim \, 5.5 \times 10^{-27} \, \left(\frac{1 \,{\rm MHz}}{f^{\rm peak}_{\rm GW}}\right) \, \log\left[2.0 \times 10^{18} \, \left(\frac{1 \,{\rm MHz}}{f^{\rm peak}_{\rm GW}}\right)^{2} \, \left(\frac{\gamma_{CP}}{10^{-2}}\right)^{2} \, \left(\frac{8.7 \times 10^{-11}}{Y_B}\right)^{2} \right]^{-1},
\ee
for 
\be \label{mxcomp1a}
5.41 \times 10^2 \left(\frac{f^{\rm peak}_{\rm GW}}{1\,{\rm MHz}}\right)^3\,\,{\rm GeV} \lesssim m_X \lesssim 1.33 \times 10^8 \left(\frac{f^{\rm peak}_{\rm GW}}{1\,{\rm MHz}}\right)^2 \,\,{\rm GeV},
\ee
and
\be \label{strainscaling2a}
h_c \, \sim \, 5.5 \times 10^{-27} \, \left(\frac{1 \,{\rm MHz}}{f^{\rm peak}_{\rm GW}}\right) \, \log\left[ 5.9 \times 10^{14} \, \left(\frac{f^{\rm peak}_{\rm GW}}{1 \,{\rm MHz}}\right)^{6} \, \left(\frac{10^{9} \, {\rm GeV}}{m_X}\right)^{4} \, \left(\frac{\gamma_{CP}}{10^{-2}}\right)^{2} \, \left(\frac{8.7\times10^{-11}}{Y_B}\right)^2 \right]^{-1},
\ee
for 
\be \label{mxcomp2a}
1.33 \times 10^8 \left(\frac{f^{\rm peak}_{\rm GW}}{1\,{\rm MHz}}\right)^2 \,\,{\rm GeV} \lesssim m_X \lesssim 1.04 \times 10^8 \ \left(\frac{\gamma_{CP}}{10^{-2}}\right)^{1/2} \, \left(\frac{8.7\times10^{-11}}{Y_B}\right)^{1/2} \, \left(\frac{f^{\rm peak}_{\rm GW}}{1 \,{\rm MHz}}\right)^{5/2} \,\,{\rm GeV}.
\ee
We note that the lower and upper bounds in Eq.~(\ref{mxcomp1a}) come from requiring $T_{\rm eva} < m_X$ and $m_X < T_{\rm PBH}$, respectively. Since $T_{\rm eva}$ and $T_{\rm PBH}$ are obtained in terms of $M_{\rm PBH}$ from Eq.~(\ref{eq:Trevap}) and Eq.~(\ref{eq:temp}) respectively, and $M_{\rm PBH}$ can be expressed in terms of the peak GW frequency through Eq.~(\ref{eq:fGWpeak}), this allows us the express the bounds on $m_X$ in terms of $f^{\rm peak}_{\rm GW}$. Similarly, the lower and upper bounds in Eq.~(\ref{mxcomp2a}) are obtained by requiring $T_{\rm PBH} < m_X$ and $\beta < \beta_c$, respectively.

\subsection{Gravitational Waves and PBH Domination}\label{sec:GWfromMD}

The calculation of GWs induced during a period of matter (PBH) domination has been considered by several authors. We refer to \cite{Domenech:2021ztg, Inomata:2019ivs, Carr:2009jm, Harada:2016mhb, Papanikolaou:2020qtd, Bhaumik:2022pil} for details and mention only a few major features here. Firstly, the peak frequency from secondary GWs from a PBH-dominated era lies in the $\mathcal{O}({\rm kHz})$ range and is given by \cite{Domenech:2021ztg}
\be
f^{\rm peak}_{\rm GW} \, \sim \, 10^{3} \, {\rm Hz} \times \left(\frac{M_{\rm PBH}}{10^4 {\rm g}}\right)^{-5/6}\,\,.
\ee
Since PBH domination occurs for $3 \times 10^4 {\rm g} \, \lesssim \, M_{\rm PBH} \, \lesssim \, 10^9 $ g, the peak of the GW spectrum falls in the range of the Einstein Telescope (for the lighter end of the allowed PBH masses) and BBO or DECIGO (for the heavier end of the allowed PBH masses). 

Secondly, the amplitude of the induced GWs at the peak frequency has been estimated to be
\be
\Omega_{\rm GW} \, \sim \, 10^{-7} \times \left(\frac{\beta}{10^{-6}}\right)^{16/3} \left(\frac{M_{\rm PBH}}{10^4 {\rm g}}\right)^{34/9},
\ee
for very sharp mass functions, and a correspondingly suppressed amplitude for wider ones. Further exploration of GWs in an era of PBH domination is beyond the scope of the current work. We therefore delineate all  regions of parameter space where PBH domination occurs and reserve them for future study.

\section{Double-Peaked Gravitational Waves and Dark Matter-Baryogenesis Coincidence Problem}
\label{sec:CoincidenceProblem}

 Probes of DM cover  numerous possible candidates and  production mechanisms and span over many  decades of possible DM masses and couplings, with the aim of explaining  the observed relic abundance $\Omega_\text{CDM}h^2 = 0.1200 \pm 0.0012$ \cite{Planck:2018nkj}. 
Given that the production of DM in the early Universe is totally  independent of baryogenesis, and the models employed to address the two questions are usually very different and independent of each other, it is a mystery why  their abundances in the current Universe come out so close to each other: $\Omega_\text{CDM}/\Omega_B \sim 5$. This is the so-called coincidence problem and has led to a huge amount of work in  the community. Particularly well-motivated proposals that solve the coincidence problem include asymmetric DM \cite{Zurek:2013wia}, cladogenesis and its variations \cite{Allahverdi:2010rh,  Allahverdi:2013tca, Kane:2011ih}, pangenesis \cite{Bell:2011tn}, cogenesis \cite{March-Russell:2011ang, Kamada:2012ht}, hylogenesis \cite{Davoudiasl:2010am}, etc. 

PBHs have been utilized in the context of the coincidence problem. One option is to  rely on Hawking evaporation of PBHs as a source of both baryogenesis and DM production \cite{Smyth:2021lkn};  this is reminiscent of models like cladogenesis or hylogenesis, which rely on the decay of a heavy beyond-SM species such as a modulus to give rise to DM and baryons. 
Other ideas of addressing the coincidence problem with PBHs have also been proposed recently \cite{ Wu:2021gtd, Balaji:2022rsy, Garcia-Bellido:2019vlf, 
Carr:2019hud}. The general tenor of these forays is to invoke a  \textit{single} population of  PBHs to address DM, baryogenesis, or both. On the other hand,  it is in some sense natural to ask whether \textit{multiple}  disparate populations of PBHs can play a role in addressing different (ostensibly correlated) cosmological phenomena. From a UV perspective, two or more phases of PBH production  may be deemed ``fine-tuned",  although typically the inflaton potential has enough freedom  to afford multiple phases of  ultra slow roll\footnote{For fine-tuning estimates in achieving PBH formation, we refer to \cite{Hertzberg:2017dkh}. At the level of a canonical single-field inflationary potential, a single phase of PBH production requires approximately one-part-in-million fine-tuning of the coefficients of the potential.}. The issue of fine-tuning notwithstanding, multiple PBH populations would yield spectacular gravitational wave (GW) signals, in the form of correlated multiple peaks.

In this section, we outline a solution to  the DM-baryogenesis coincidence problem by invoking \textit{two} populations of PBHs: a population of light primordial black holes (LPBH) that is responsible for baryogenesis, and a population of heavy primordial black holes (HPBH) that constitutes DM. The LPBHs induce baryogenesis by the method described in the previous sections.
The DM-baryon coincidence in our scenario is obtained as a ratio of the abundances and masses of the LPBH and HPBH populations,  the number of relativistic degrees of freedom at the relevant times, and, from the particle physics side, the baryon asymmetry produced per decay of $X$. The main observational  signature of our scenario  lies in the  GW spectrum, since curvature perturbations responsible for PBH formation also generate GW signals. The GW spectrum will consist of two peaks corresponding to the two populations of PBHs. For the population of heavy PBHs to be stable today in order to constitute all of DM, we need $M_{\rm HPBH} \gtrsim 10^{15}$ g and the corresponding peak frequency is $\sim \mathcal{O}({\rm Hz})$. In frequency space, this translates to a hierarchy of $\gtrsim 10^6$ Hz between the two peaks corresponding to baryogenesis and DM.

While we reserve a full study of the relevant UV physics of PBH formation for the future, we note that the main feature is that the PBH mass function is required to be bimodal. Generally speaking, if PBHs are required to play some \textit{combination} of roles in cosmology, then  a multimodal mass function becomes a necessity, and our model is a special case of the scenarios envisioned in  \cite{Carr:2018poi}\footnote{Multimodal PBH mass functions have been investigated in other contexts by several groups recently. They may arise from oscillations \cite{Cai:2018tuh, Cai:2019jah} or  multi-parametric resonances \cite{Addazi:2022ukh} in the sound speed during inflation, multiple stages of ultra slow-roll inflation \cite{Zhang:2021vak}, or oscillatory features in the power spectrum of inflation from a choice of non-Bunch-Davies vacua \cite{Carr:2018poi, Armendariz-Picon:2003knj}.}.

\subsection{Bi-modal mass function}
\label{csec:PBHmassfunction}

We begin our discussion by appropriately modifying the PBH mass function. We assume that there are two mass function peaks and define
\bea
\frac{d\beta(M_{\rm PBH},t_i,t)}{d\log M_{\rm PBH}} = \frac{d\beta_L(M_{\rm L},t_{i,L},t)}{d\log M_{\rm L}}+\frac{d\beta_H(M_{\rm H},t_{i,H},t)}{d\log M_{\rm H}},
\eea
with the subscripts ``$L$" and ``$H$" standing, respectively, for the population of LPBHs and HPBHs. We then have 
\begin{equation}
    \beta_L=\frac{n_L(t_{i,L})M_L}{\rho(t_{i,L})},~~~~~ \beta_H=\frac{n_H(t_{i,H})M_H}{\rho(t_{i,H})}.
\end{equation}
We parameterize the double peaked power spectrum with peak locations at  $k_0=k_{\rm H/L}$ as
\bea
P_{\zeta}(k)&=&P_{\zeta,H}(k)+P_{\zeta,L}(k) \nonumber \\
&=&A_{H}\delta\left(\log \left(\frac{k}{k_H}\right)\right) + A_{L}\delta\left(\log \left(\frac{k}{k_L}\right)\right).\label{doubledelta2}
\eea
Here,  $P_{\zeta,H}$ is responsible for the production of the heavy PBHs and $P_{\zeta,L}$ for the light PBHs.

\subsection{Coincidence Problem}
If  HPBH formation happens within the radiation-dominated  era, the contribution of HPBHs to the cold DM abundance today, which is expressed in terms of the ratio of the current HPBH mass density to that of the CDM density, 
\begin{equation}
    f=\frac{\Omega_\text{HPBH}}{\Omega_\text{CDM}},
\end{equation}
can be evaluated as follows~\cite{Carr:2009jm}. In the remainder of this section, we call the masses of the two populations of PBHs as $M_L$ and $M_H$, standing for light PBHs responsible for baryogenesis and HPBHs responsible for DM, respectively. Similarly, the abundances of the  two species are labelled $\beta_L$ and $\beta_H$.

The fraction of the energy of Universe in HPBHs at their formation time is related to their number density during the radiation era by 
\begin{equation}
    \beta_H=\frac{\rho_\text{HPBH}(t_i)}{\rho_\text{R}(t_i)}=\frac{M_H n_H(t_{i,H})}{\rho_\text{R}(t_{i,H})}=\frac{4}{3}\frac{g_{\star,S}(T_{i,H})}{g_{\star}(T_{i,H})}\frac{M_H}{T_{i,H}}\frac{n_H(t_{i,H})}{s(t_{i,H})}=\frac{4}{3}\frac{g_{\star,S}(T_{i,H})}{g_{\star}(T_{i,H})}\frac{M_H}{T_{i,H}}\frac{n_H(t_0)}{s(t_0)}.
\end{equation}
Therefore the current density parameter for
HPBHs is given by
\begin{eqnarray}
 \Omega_\text{HPBH}=\frac{M_H n_H(t_0)}{\rho_c}&=&\frac{3}{4}\frac{g_{\star}(T_{i,H})}{g_{\star,S}(T_{i,H})}\beta_H\frac{s(t_0)}{\rho_c}T_{i,H}(M_H)\\
 &=&\left(\frac{2.7\times 10^8}{\text{GeV}}\right)h^{-2}\frac{3}{4}\frac{g_{\star}(T_{i,H})}{g_{\star,S}(T_{i,H})}\beta_H T_{i,H}(M_H),
\end{eqnarray}
since $s(t_0)=2891.2\,\text{cm}^{-3}$ and $\rho_c=1.052\times10^{-5}h^2\,\text{GeV}\,\text{cm}^{-3}$.

Hence
\begin{eqnarray}
\nonumber f=\frac{\Omega_\text{HPBH}}{\Omega_\text{CDM}}&=&\left(\frac{2.7\times 10^8}{\text{GeV}}\right)\left(\frac{1}{\Omega_\text{CDM} h^2}\right)\frac{3}{4}\frac{g_{\star}(T_{i,H})}{g_{\star,S}(T_{i,H})}\beta_H T_{i,H}(M_H)\\
&=&\left(\frac{1}{\Omega_\text{CDM} h^2}\right)\frac{3}{4}\beta_H \left(\frac{T_{i,H}(M_H)}{6.6 \times10^{-33}\text{g}}\right).
\end{eqnarray}

The initial abundance of HPBHs which can explain a fraction $f$ of cold DM today is given by
\begin{equation}
    \beta_H=\frac{8\pi^{3/4}}{3\sqrt{3}\,5^{1/4}}\frac{g_{\star,S}(T_{i,H})}{g^{3/4}_\star(T_{i,H})}\frac{\rho_c}{s(t_0)}\gamma^{-1/2}\Omega_\text{CDM}f\left(\frac{M_H}{M_\text{Pl}^3}\right)^{1/2}.
    \label{eq:betahlimit}
\end{equation}

To understand the coincidence problem in the context of this scenario, we obtain the ratio $f/Y_B$:
\begin{eqnarray}
  \nonumber \frac{f}{Y_B}&=&\frac{\beta_H}{\beta_L}\frac{1}{\gamma_{CP}}\frac{1}{ \Omega_\text{CDM} h^2}\frac{1}{N_\chi}\frac{T_{i,H}(M_H)}{T_{i,L}(M_L)}\left(\frac{M_L}{6.6 \times10^{-33}\text{g}}\right)\\
  &=&\frac{\beta_H}{\beta_L}\frac{1}{\gamma_{CP}}\frac{1}{\Omega_\text{CDM} h^2}\frac{1}{N_\chi}\left(\frac{g_\star(T_{i,L})}{g_\star(T_{i,H})}\right)^{1/4}\sqrt{\frac{M_L}{M_H}}\left(\frac{M_L}{6.6 \times10^{-33}\text{g}}\right)\,\,. \label{Bdmratio}
\end{eqnarray}
By fixing the particle physics ($m_X$ and $\gamma_{CP}$), the coincidence problem thus manifests itself as the right choice of PBH masses, $M_L$ and $M_H$, and their initial abundances, $\beta_L$ and $\beta_H$. 

The left panel of Fig.~\ref{fig:MD2} shows the upper limit on the initial abundance of HPBHs that can explain the whole abundance of DM today, as a function of the mass of HPBHs. The lower limit on the mass of HPBHs, $10^{15}\,\text{g}$ corresponds to the minimum mass of HPBHs that can survive until today.  We note that HPBHs are subject to other cosmological and astrophysical constraints that we did not show in this Figure. For an extensive review of these constraints and their current status see~\cite{Carr:2016drx, Carr:2020gox, Carr:2020xqk, Green:2020jor, Villanueva-Domingo:2021spv} and references within.

\subsection{Formation of Heavy Primordial Black Holes in a Matter- (PBH-) Dominated Era}
The possibility of formation of PBHs within an early matter-dominated era was first studied as a cosmological probe of grand unified theories~\cite{Khlopov:1980mg,1982SvA....26..391P}. In these models, the predicted unstable heavy particles can dominate the energy density of the Universe before their decay and start an early matter-dominated epoch. While the lack of pressure enhances the probability of formation of PBHs within an early matter-dominated era, inhomogeneity and anisotropy of density perturbations can hinder formation of PBHs~\cite{Khlopov:1980mg,1982SvA....26..391P,Harada:2016mhb}. It can be shown that 
inhomogeneity and anisotropy of density perturbations reduce the probability of formation by a factor of $\sigma^{3/2}$ and $\sigma^{5}$ respectively. The probability of formation of PBHs during an early matter-dominated era is shown to be equal to~\cite{Khlopov:1980mg}:
\begin{equation}
    \beta(M_\text{PBH}) \approx 0.02\sigma^{13/2}(M_\text{PBH}).
    \label{eq:beta1}
\end{equation}
If the early matter-dominated epoch begins at $t=t_{\text{m}, i}$ and ends at $t=t_{\text{m}, f}$, then the mass of PBHs form during this time interval falls in the following mass range~\cite{Khlopov:1980mg}:
\begin{equation}
    M_\text{min}\sim M_\text{hor}(t_{\text{m}, i})\lesssim  M_\text{PBH}\lesssim M_\text{max}\sim M_\text{hor}(t_{\text{m}, f})\sigma(M_\text{max})^{3/2},
    \label{eq:massrangeMD}
\end{equation}
where $M_\text{hor}$ denotes the horizon mass, and $M_\text{max}$
is obtained by assuming that fluctuations at scale $M_\text{max}$ enter the horizon at $t(M_\text{max})$ and grow to nonlinear regime and decouple before $t_{\text{m}, f}$~\cite{Khlopov:1980mg}.

\begin{figure}[t]
  \centering
    \includegraphics[width=0.4\textwidth]{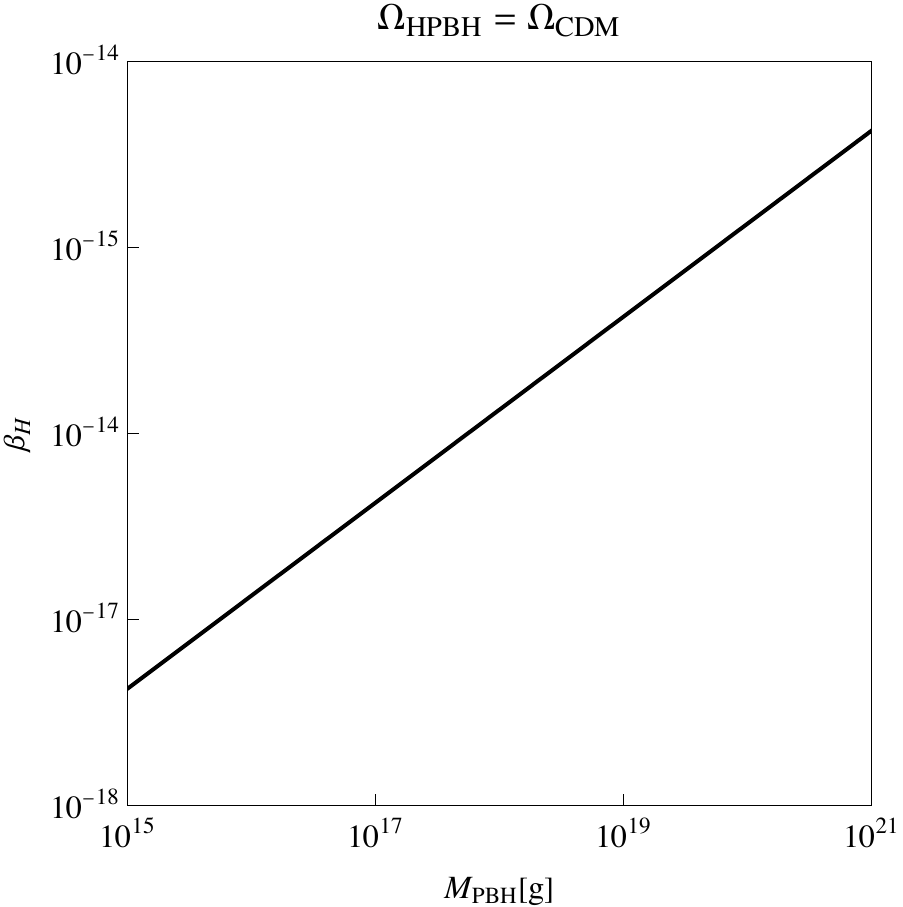}\quad
    \includegraphics[width=0.373\textwidth]{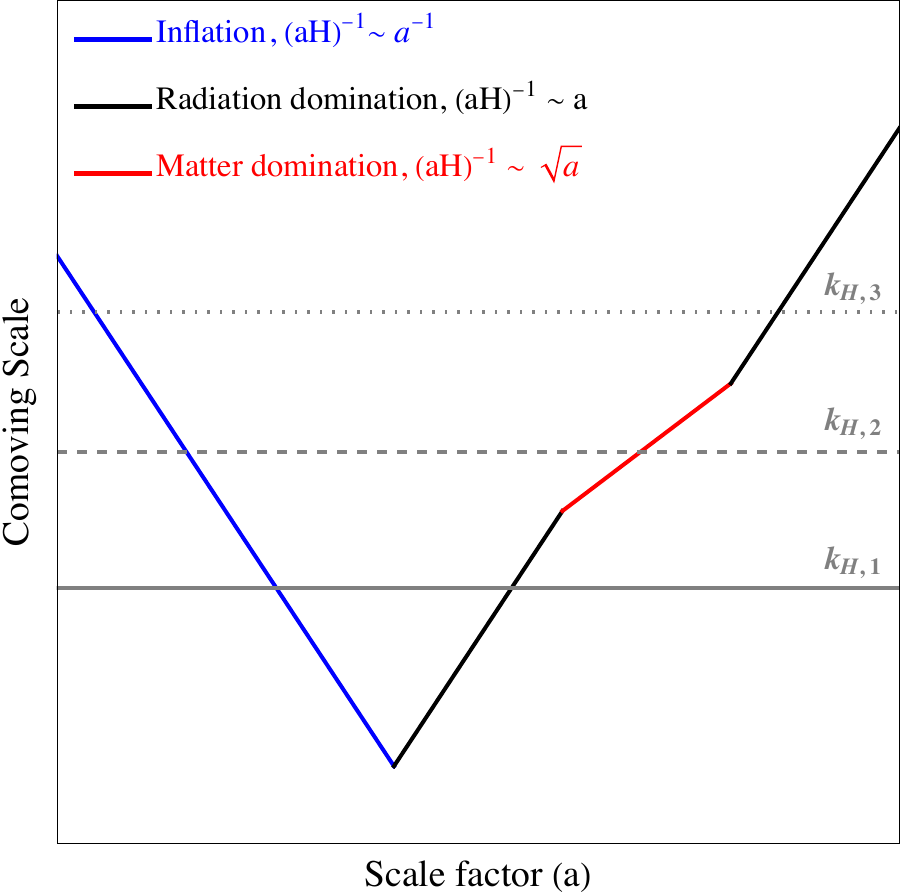}
  \caption{\textit{Left panel}: The initial abundance of HPBHs, $\beta_H$, as a function of the HPBH mass, assuming that HPBHs can explain all the observed abundance of DM. \textit{Right panel}: The different options for the formation of HPBHs are displayed. The segment in red denotes a phase of LPBH (matter) domination. $k_{H,1}$, $k_{H,2}$, and $k_{H,3}$, denote, respectively, the cases when HPBHs are formed prior to, during, and after the LPBH dominate the energy density of the Universe.
  }
  \label{fig:MD2}
\end{figure}

A population of HPBHs can form when a population of LPBHs dominate the energy density of the Universe, if the heavy mode, $k_H$, which is responsible for formation of HPBHs, re-enters the horizon during LPBH domination. 
We assume LPBHs form at $t=t_{i,L}$ with probability $\beta_L$ and with a mass equal to $M_L=\gamma M_\text{hor}(t_{i,L})$ in a radiation-dominated Universe, they dominate the energy density at $t=t_{L\text{D}}$, and eventually they evaporate at $t=t_{L,\text{eva}}=t_{i,L}+\tau_L\simeq\tau_L$, where $t_{i,L}\lesssim t_{L\text{D}}\lesssim \tau_{L}$. To guarantee that LPBHs come to dominate the energy density of the Universe before their evaporation, we require $\beta_L\geq\beta_c$. We denote the time of re-entry of $k_H$ mode to horizon and equivalently the formation time of HPBHs by $t=t_{i,H}$ where $t_{L\text{D}}\lesssim t_{i,H}\lesssim \tau_{L}$ to make sure that formation happens within LPBHs domination.
According to Eq.~(\ref{eq:massrangeMD}), production of HPBHs enhances over the mass range given by:
\begin{equation}
     M_\text{hor}(t_{L\text{D}})\lesssim  M_{H}\lesssim M_\text{hor}(\tau_L)\sigma^{3/2}(M_\text{Max}).
\end{equation}
The initial abundance of LPBHs, $\beta_L$, selects the domination time, $t_{L\text{D}}$, from the range $t_{i,L}\lesssim t_{L\text{D}}\lesssim \tau_{L}$. For $0.5\leq\beta_L\leq1$, LPBHs domination happens at their formation time, i.e.,  $t_{L\text{D}}=t_{i,L}$, and this corresponds to the smallest possible HPBH mass. For $\beta_c\leq \beta_L<0.5$, LPBHs domination happens before their evaporation. The maximum possible mass for HPBHs is determined by $\tau_L$ via Friedemann equation, $H(\tau_L)=2/(3\tau_L)$. Therefore, for $\beta_L\geq\beta_c$, if $k_H$ re-enters the horizon while LPBHs dominate the Universe, HPBHs can form within the following mas range:
\begin{equation}
     M_L/\gamma\lesssim  M_{H}\lesssim \frac{7680\pi}{g_\star(T_\text{LPBH})}\frac{M_L^3}{M_\text{Pl}^2}\sigma^{3/2}(M_\text{Max}).
\end{equation}
In Fig.~\ref{fig:MD2} (right panel), we display  the different possibilities for formation of HPBH. The heavy mode can re-enter the horizon within a radiation-dominated era succeeded by an early matter-(LPBH-) dominated era  ($k_{H,1}$), or during an early matter-dominated era followed by a secondary radiation-dominated epoch  ($k_{H,2}$), or in a radiation-dominated era proceeded by an early matter-dominated era ($k_{H,3}$).
The treatment for  HPBHs formed during an era of LPBHs domination is beyond the scope of this paper and is left for future work.

\subsection{Double-Peaked Gravitational Waves}

\begin{figure}[h]
\centering
\hspace{2cm}
\includegraphics[scale=0.5]{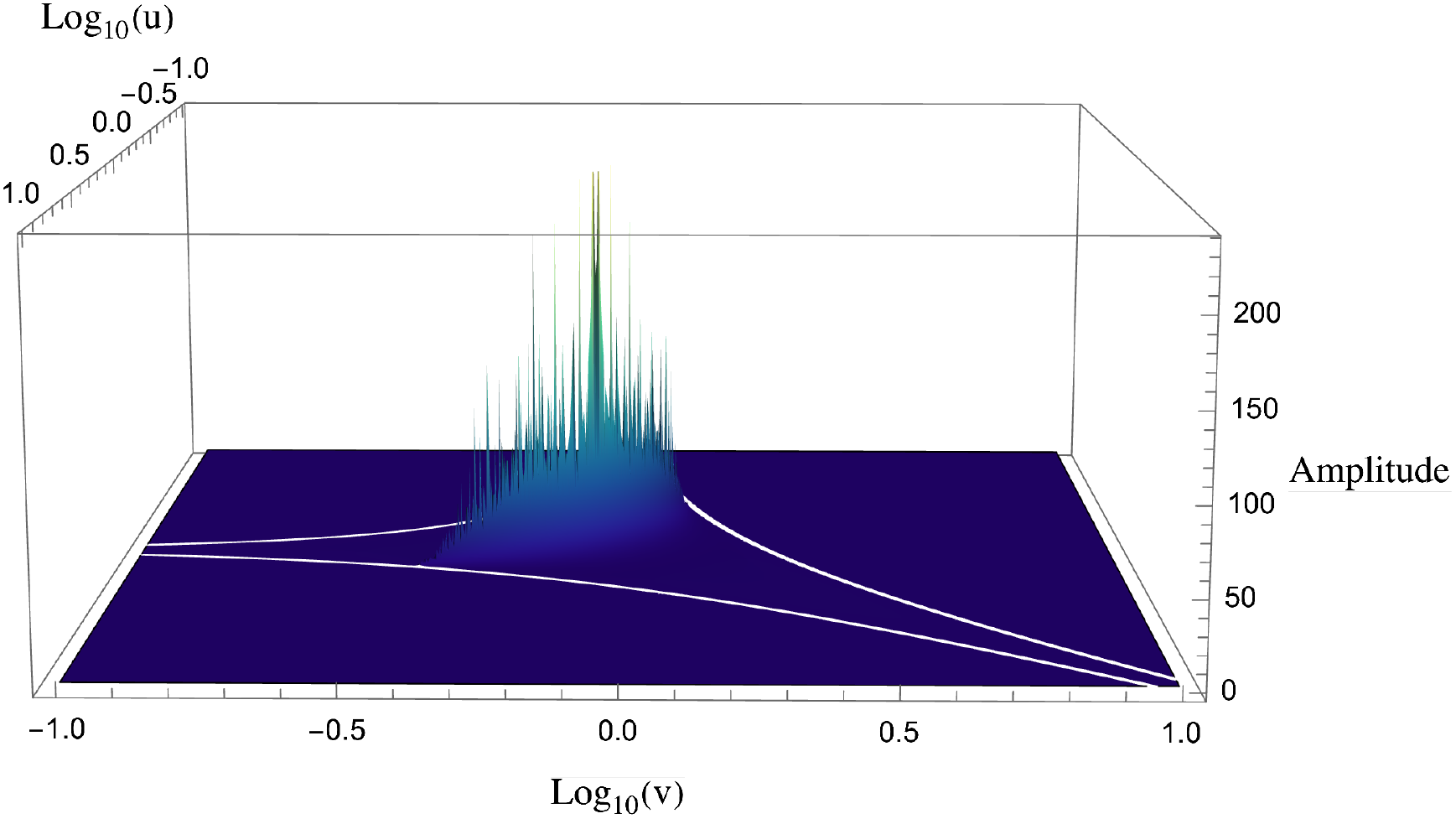}
\caption{The relative amplitude of cross term contribution to $P_h$ as a function of wave numbers of two scalar modes.}
\label{fig:crossingterm1}
\end{figure}

In this subsection, we provide a computation of the GW spectrum coming from the two populations  of  PBHs. As noted in Eq.~(\ref{doubledelta2}), the power spectrum consists of two contributions: $P_{\zeta,H}$, which is responsible for the production of the heavy PBHs and $P_{\zeta,L}$, which is responsible for the light PBHs. We first turn to the question of whether there could be an additional contribution to the GW signal from the crossing term $P_{\zeta,H} \times P_{\zeta,L}$. 
 In Fig.~\ref{fig:crossingterm1}, we show the relative amplitude of the cross term contribution to $P_h$ with arbitrary choice of two scalar modes\footnote{Here we ignore the decay of sub-horizon modes in order to illustrate the property of the tensor mode. The decay of sub-horizon modes during radiation domination will further suppress the GW signal.}. The $u/v$ parameter is the ratio of the chosen tensor mode wave number $k_{\rm GW}$ and the scalar modes wave number $k_H/k_L$. As is shown in the figure, the GW amplitude is non-vanishing only in the regions with $u \sim v \sim 1$, which is equivalent to $k_{\rm GW} \sim k_H \sim k_L$. The suppression on GW production outside this region is a result of the hierarchical splitting of the two scalar modes. Momentum conservation is satisfied only when the scalar modes are close in  Fourier space. Since the scalar modes of interest in our case have a large hierarchy $k_L / k_H \sim 10^8$, the cross term contribution is negligible in our study. We leave the detailed study of GWs from cross terms to future work, since in certain cases where the hierarchy is smaller  there could be interesting effects.

\begin{figure}[h]
\centering
\includegraphics[scale=0.55]{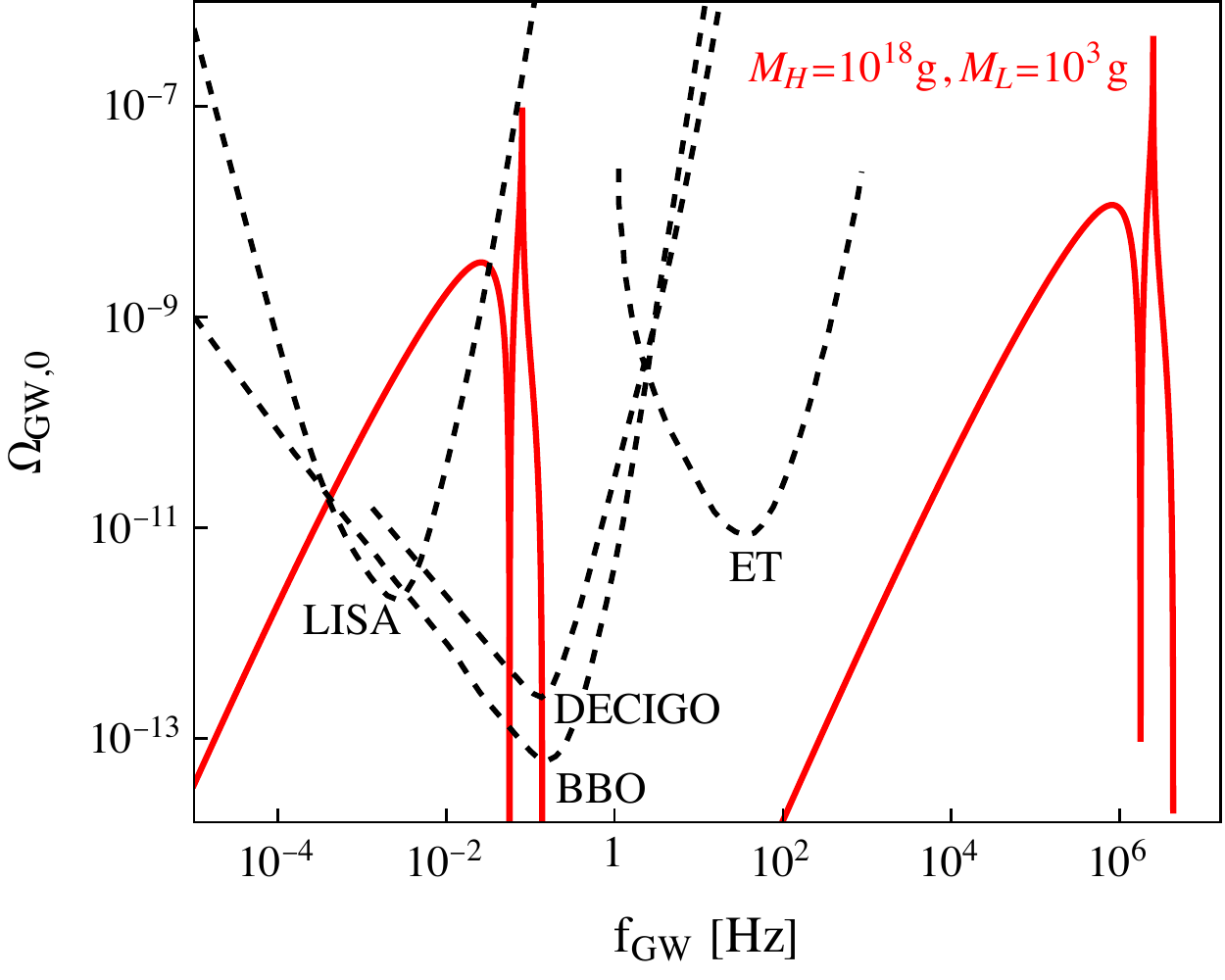}\qquad
\includegraphics[scale=0.55]{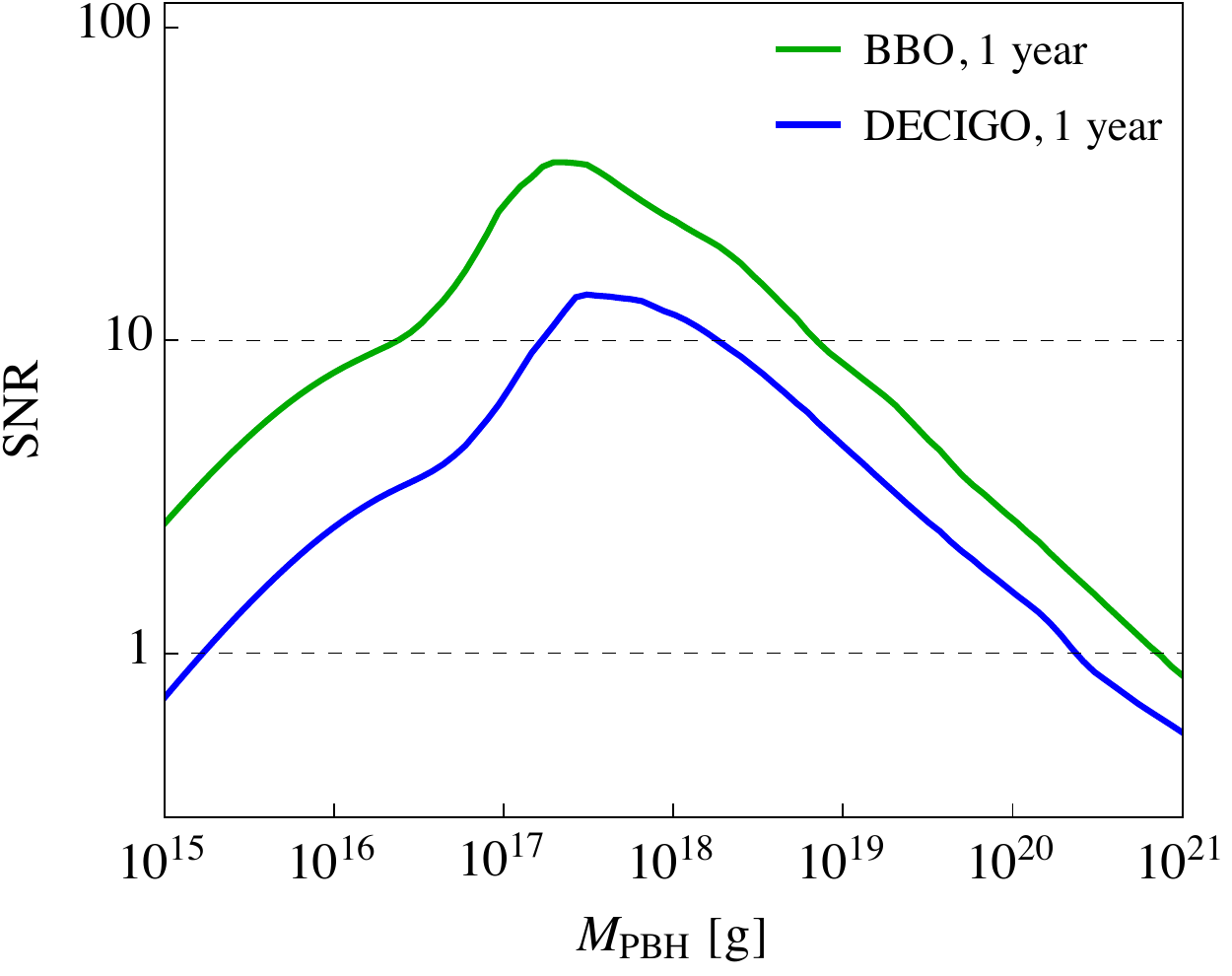}
\caption{\textit{Left panel}: A benchmark GW spectrum from  scalar perturbation modes. In the sub-Hz region, we show the GW radiation from perturbations that generated heavy PBH distributions peaked at $M_H=10^{18}$ g. In the high-frequency region, we show the GW radiation from perturbations generating lighter PBHs at $M_L= 10^{3}~{\rm g}$ with initial abundance calculated assuming $m_X<T_{\rm BH}$. Sensitivities of future GW observatories are shown with black dashed curves. \textit{Right panel}: The Signal-to-Noise Ratio of second order GW signals at the low frequency peak for future detectors like  BBO~(green) and DECIGO~(blue). The observation time is assumed to be one year. The $M_{\rm PBH}$ is the peak mass of the heavy PBH species that make up $100\%$ DM relic abundance.
}
\label{fig:GW2Peaks}
\end{figure}

Having established that the cross term contribution is negligible in our case, we calculate the full GW spectrum signal by adding contributions calculated with each monochromatic curvature perturbation peak separately. In Fig.~\ref{fig:GW2Peaks} (left panel), we show the double peaked GW spectrum that results from our model. We choose a  benchmark value of the scalar perturbation wavenumber that gives the peak location of heavy PBH mass function at $M_H=10^{18}~{\rm g}$. For the light PBHs, we also choose benchmarks of high-$k $ modes that gives the peak location $M_L=10^{3}~{\rm g}$. We further require $f=1$ and $Y_B = 8.7 \times 10^{-11}$ in Eq.~(\ref{Bdmratio}). Since the horizon re-entry time is well defined by the horizon mass of interest, the peak location of the GW spectrum can be estimated directly from the mass of the light and heavy PBH species. Therefore the bimodal PBH mass function is strongly correlated with a double-peaked GW spectrum.

The GW signal from the formation of heavy PBHs peaks at $f_{\rm GW}\sim~{\rm Hz}$. Future GW observatories, for example BBO~\cite{Corbin:2005ny} and DECIGO~\cite{Seto:2001qf,Kawamura:2020pcg}, can perform a precise measurement of the scalar perturbations in this region. In Fig.~\ref{fig:GW2Peaks} (right panel), we show the Signal-to-Noise Ratio (SNR) of the induced GW signals from $P_{\zeta,H}$ with one-year measurements in BBO and DECIGO. The peak location $M_{\rm PBH}$ of the heavy PBH mass function is calculated with the horizon mass in the course of $k_H$ re-entry in Eq.(\ref{eq:MR}). The amplitude of $P_{\zeta,H}$ is chosen such that heavy PBHs explain the DM relic abundance in the cosmological coincidence problem. As is discussed previously, the interference  between the two peaks can be neglected  because of the large separation in their frequencies. We calculated the GW signal in the BBO and DECIGO frequancy band with only $P_{\zeta,H}$ and checked the crossing term contribution is negligible. As a result of the large density perturbation for PBH formation, the SNRs in both detectors are larger than one for the mass region that PBH can make up DM. 

The main observational feature here is the double-peaked GW signal. In  Section~\ref{UVcomp} of the Appendix, we give an outline for constructing an inflaton  potential that can give rise to two  spikes in the power spectrum, through two ultra slow-roll phases.

\section{CONCLUSIONS}
\label{sec:conclusion}

The purpose of this paper has been to highlight the importance of the ultra high frequency GW frontier for topics like baryogenesis that have traditionally been of interest to particle theorists. Broadly speaking, electroweak baryogenesis is the main setting where the confluence of collider and GW physics has been extensively studied (we refer to some representative papers \cite{Ramsey-Musolf:2019lsf, Profumo:2007wc, Caldwell:2022qsj, Alves:2018oct, Alves:2018jsw, Huang:2016cjm} in this vast  literature). This is of course driven by the classic status of electroweak baryogenesis, but also by the fact that  the peak frequency of the GWs in this case is in the target window of upcoming space-based detectors. In contrast, the high frequency regime where PBH-induced baryogenesis can occur is much more challenging to probe, but, in our opinion, no less important. PBHs, if they exist, provide a window into the very early inflationary era of the Universe and also connect to topics like baryogenesis, DM, and dark radiation. We have calculated the parameter space on the strain versus peak frequency plane for baryogenesis to successfully occur, being careful to identify washout and PBH-dominated regimes. The results have been shown for different values of the particle physics inputs, such as $m_X$ and $\gamma_{CP}$. Semi-analytic scaling have been provided. We have also outlined a scenario where two populations of PBHs: one heavy and one light - can address the DM-baryogenesis coincidence problem. 

There are several future directions to explore. Firstly, in this study, we have focused on the scenario where the energy density of PBHs is always a sub-dominant component of the total energy density of the universe. Therefore, GW signals produced during horizon re-entry  redshift with the cosmological expansion as a fixed fraction of radiation. On the other hand, any non-standard cosmology that dilutes the energy density of radiation will also affect the GW signal, leading to a suppressed strain today. In particular, the dilution could happen when PBHs are over-produced from a large curvature perturbation such that there was a PBH-dominated epoch in the early universe.
 While there is nothing inconsistent about transitioning to a PBH-dominated phase, the GW signal is typically suppressed when such a phase occurs. Nevertheless, it would be interesting to investigate how the predicted signals in our scenario change when one also has a PBH-dominated cosmology. This is especially interesting when a population of heavy PBHs starts forming in the background of a cosmology dominated by a population of light PBHs.
 
 It is also important to probe the correlations between DM, dark radiation, PBHs, and GWs along the same lines as has been pursued in this work. The most important future direction is to creatively think about and propose detection techniques for GWs in the MHz - GHz range.

\acknowledgments
The work of B.S.E is supported in part by DOE Grant DE-SC-0002424. B.S.E would like to thank the Department of Physics and Astronomy, University of Utah,
where part of the work was completed. 
The work of K.S. and T.X. is supported in part by DOE Grant desc0009956 and T.X. is supported by the Israel Science Foundation (grant No. 1112/17).

\appendix

\section{Possible Inflaton Potentials for Multimodal PBH Distributions} \label{UVcomp}

The formation of PBHs is exponentially  sensitive  to the ratio of the density perturbation to the critical density. This implies  that modulations to the primordial power spectrum can have a drastic effect on the mass function. In particular, multimodal PBH mass functions have been discussed in three broad classes of UV settings:  $(i)$ specific features of the inflaton potential that  induce distinct phases of ultra slow-roll, such as multiple inflection points \cite{Hertzberg:2017dkh, Ozsoy:2018flq, Franciolini:2022pav, Cicoli:2018asa, Zhang:2021vak, Gao:2021lno}; $(ii)$  oscillatory features in the primordial spectrum emanating from the ambiguity in defining the vacuum state inflation \cite{Danielsson:2002kx}, \cite{Armendariz-Picon:2003knj}; and $(iii)$ oscillatory features in the speed of sound during inflation  \cite{Cai:2018tuh}. Of these, only the option of multiple ultra slow-roll phases is able to achieve multiple peaks separated by a large hierarchy in frequencies, as is required in our case.

For completeness, we briefly provide a road map for achieving multiple ultra slow-roll phases at the level of an effective description of the scalar potential. A generic potential for a scalar field $\phi$ can be characterized as
\be
V(\phi)  \, = \, V_0 \sum_{n} c_n \left( \frac{\phi}{\Lambda} \right)^n,
\ee
where $V_0$ is the overall scale  of the potential, $c_n$ are tunable coefficients, and $\Lambda$ is a cutoff scale. Requiring $j$ inflection points imposes the following conditions on $V(\phi)$:
\be
V^{\prime}(\phi_i) \, = \, 0, \,\, V^{\prime \prime}(\phi_i) \, = \, \alpha_i \sim 0,  \,\,\,\,\,\, \,\,\,\,\,\, \forall \,\,  i = 1 \ldots  j \,\,. 
\ee
Here, the  $\alpha_i$ denote values of the second derivatives of the potential at the $\phi_i$. The set $\{\phi_i, \alpha_i\}$ can be tuned to the specific values needed by the model of interest by choosing the coefficients $c_n$.

The values of $\{\phi_i, \alpha_i\}$ are set by requiring PBH formation during multiple stages of ultra slow-roll at different $k$-modes. The conditions for  PBH formation can be obtained by starting with the evolution of the background. The FRW and scalar field equations are given by
\bea
H^2 &=& \frac{1}{2}\dot{\phi}^2 + V(\phi), \nonumber \\
\ddot{\phi} &+& 3H\dot{\phi} + V^{\prime}(\phi) = 0.
\eea
Here, the dots refer to derivatives with respect to time, while the prime refers to derivatives with respect to $\phi$. The number of e-foldings is given by $dN = Hdt$.

The slow-roll parameters are given by
\be \label{slowrolldefs}
\epsilon_H \,=\, -\frac{\dot{H}}{H^2},\,\,\,\,\,\,\, \,\,\, 
\eta_H \,=\, - \frac{\ddot{H}}{2H\dot{H}}.
\ee
The  power spectrum is given by
\be \label{scpspec}
\mathcal{P}_{\mathcal{R}}(k) \, = \, \frac{1}{8\pi^2}\frac{H^2}{\epsilon_H}.
\ee
We note that Eq.~(\ref{scpspec}) underestimates the power spectrum and to obtain the precise value, one needs to numerically solve the Mukhanov-Sasaki equations. However, Eq.~(\ref{scpspec}) is sufficient to understand the qualitative behavior of the power spectrum. The requirement for PBH formation with masses $M_{\rm LPBH} = 10^3$g and $M_{\rm HPBH} = 10^{18}$g is 
\be \label{condspec}
\mathcal{P}_{\mathcal{R}}(k) \,  \gtrsim 10^{-2} \,\,\,\,\, {\rm at} \,\,\,\,\, \log_{10} k\,\sim\,\{14.1, 21.6\}\,\,.
\ee
Eq.~(\ref{condspec}) is easily satisfied for choices of $\{\phi_1, \phi_2, \alpha_1, \alpha_2\}$, which in turn can be attained by choices of $c_{2-6}$. One option is to perform a scan at the level of the coefficients $c_{2-6}$ to obtain the dynamics one needs.

A more structured algorithm has been advanced recently in the form of reverse engineering the inflaton potential given the shape of the power spectrum \cite{Hertzberg:2017dkh, Franciolini:2022pav}.  In this approach, one would start from the conditions in Eq.~(\ref{condspec}), and work out the evolution of $H(k)$,  $\epsilon_H(k)$ and $\eta_H(k)$, ultimately trading these dependencies to obtain $V(\phi)$. This can be done either working with functions of the comoving wave number $k$ or the number of e-foldings, in which case Eq.~(\ref{condspec}) roughly translates to the condition  $\mathcal{P}_{\mathcal{R}}(N) \sim 10^{-2}$ at $N \sim 33$ and $N \sim 49$. We leave a full reconstruction and study of the inflaton potential corresponding to our scenario for future work.

 \section{GW Spectrum for a  Smoother Power Spectrum} \label{smoothpsepc}

\begin{figure}[h]
\centering
\includegraphics[scale=0.58]{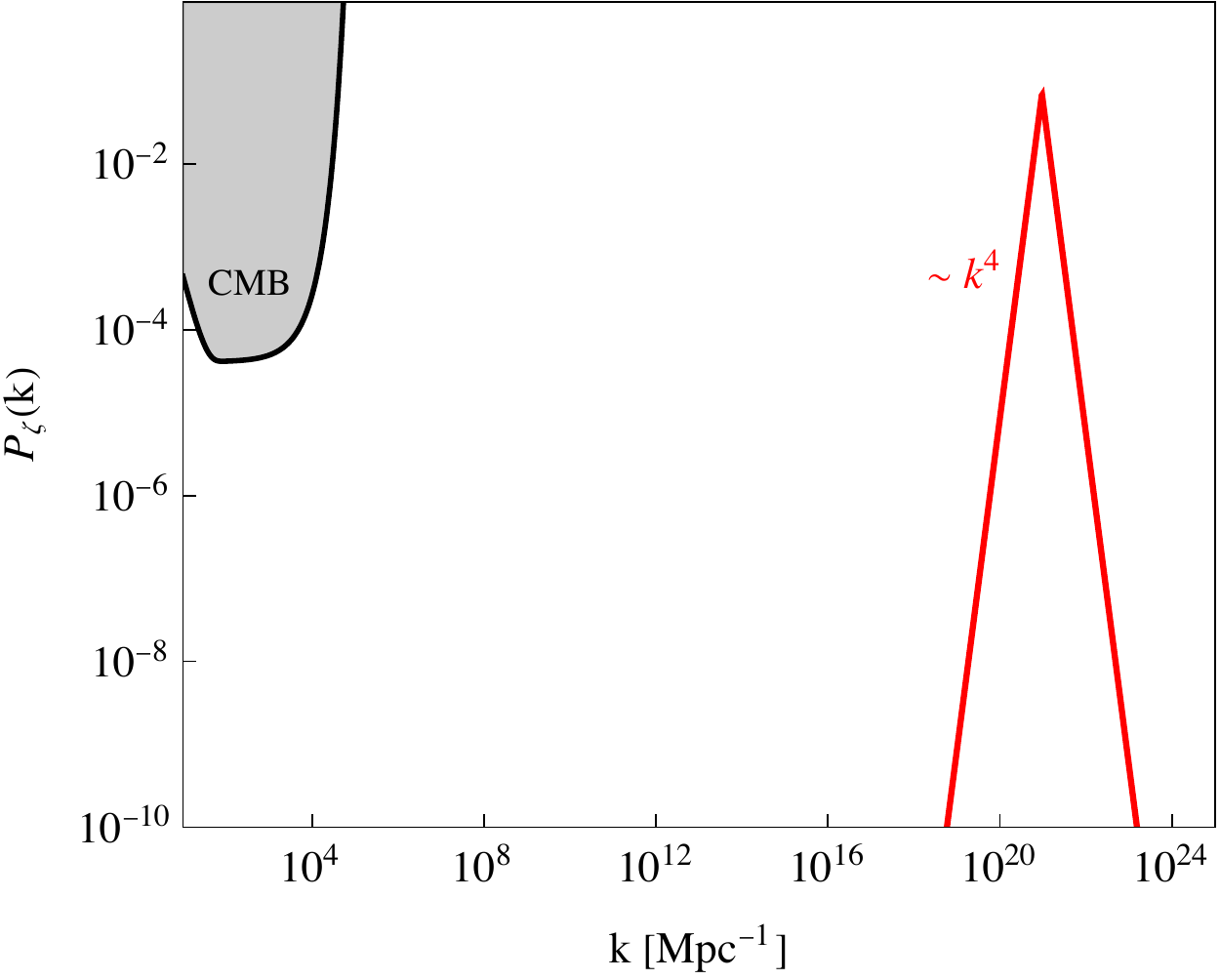}\quad
\includegraphics[scale=0.6]{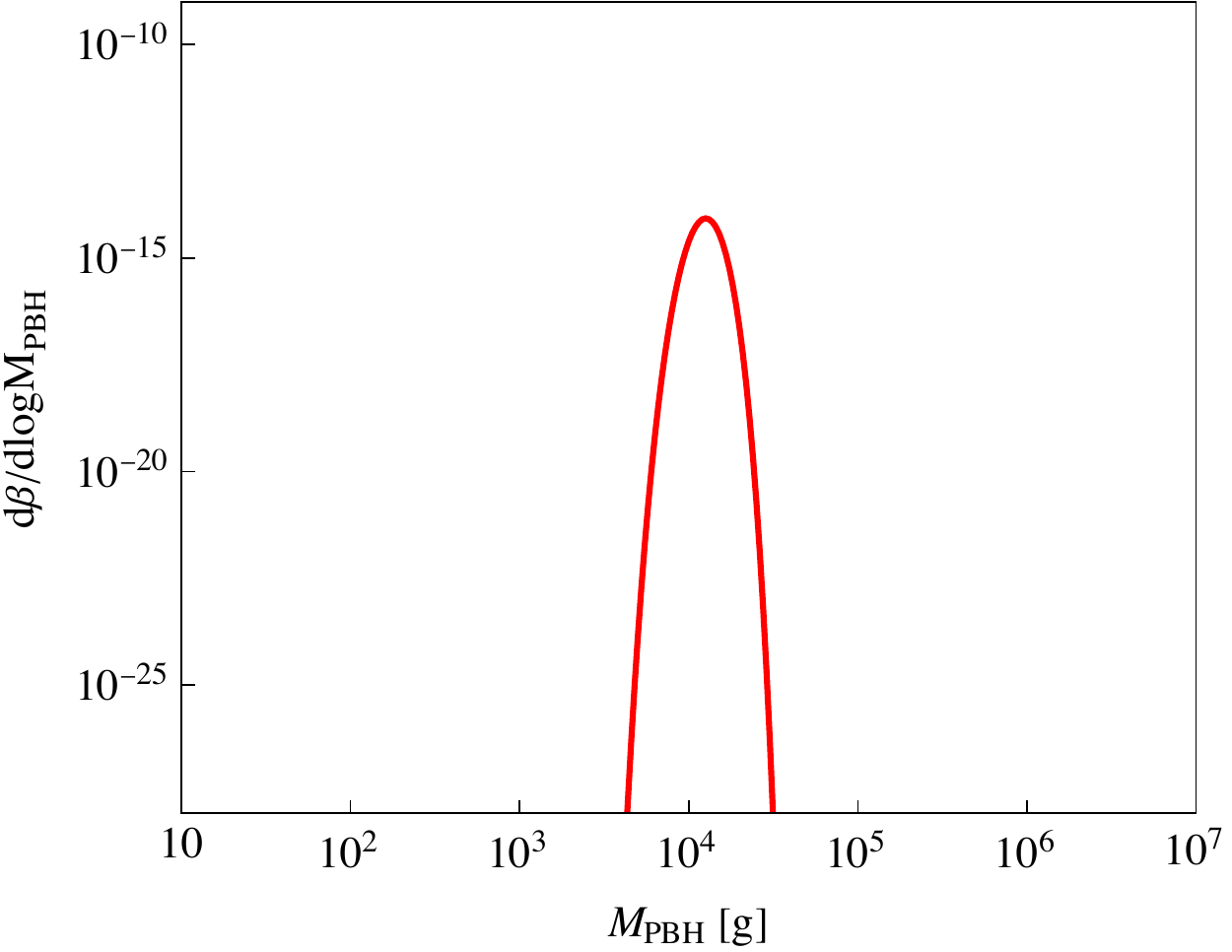}\\
\includegraphics[scale=0.59]{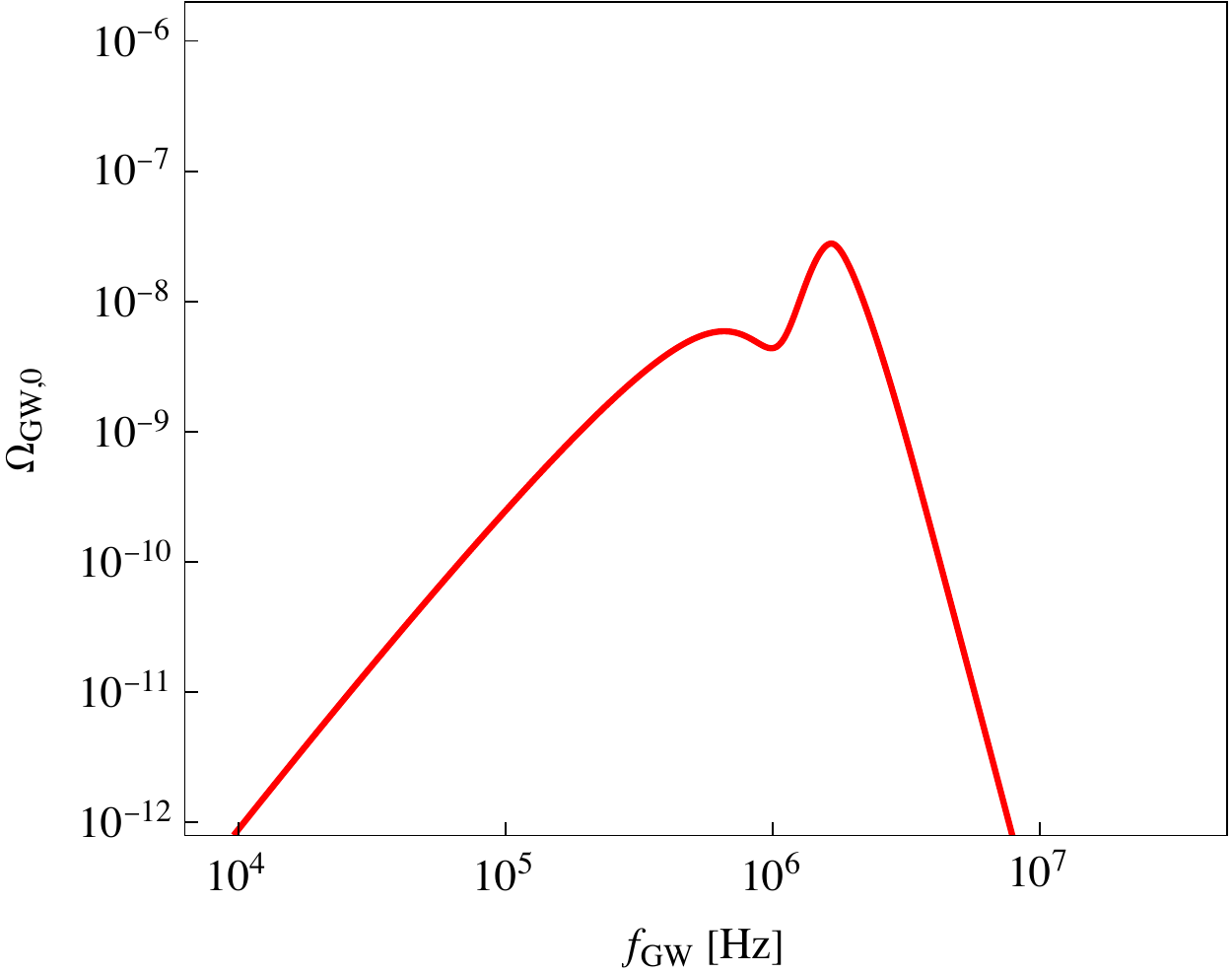}\quad
\includegraphics[scale=0.58]{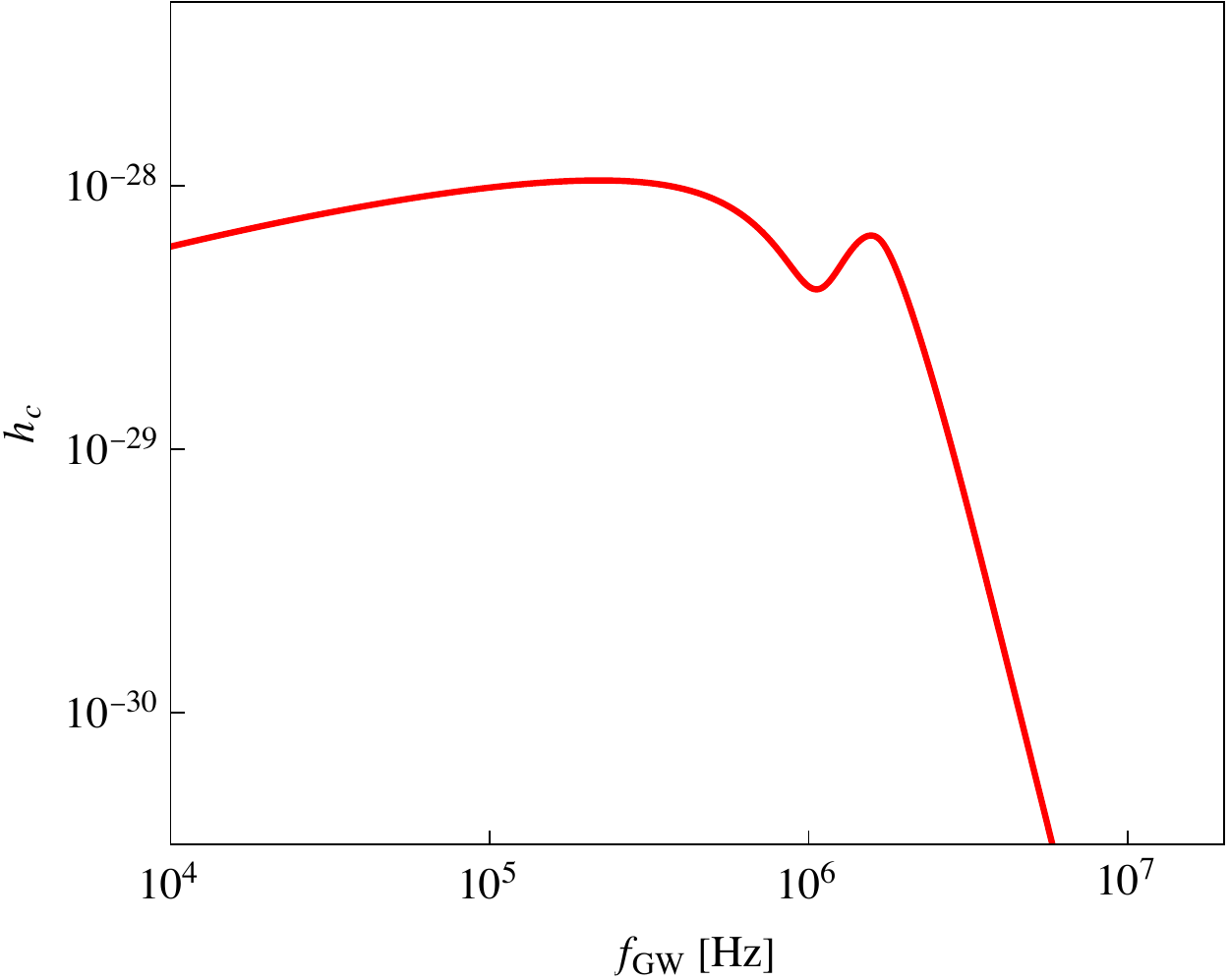}
\caption{\textit{Top-left panel}: Example power spectrum with $k^4$ scaling and peaks located at $k = 10^{21}$ ${\rm Mpc}^{-1}$ corresponding to $f^{\rm peak}_{\rm GW} = 1.5~{\rm MHz}$. The corresponding PBH masses is $M_{\rm PBH} = 10^4~{\rm g}$. The amplitude of the power spectrum is chosen for the abundance of PBHs. \textit{Top-right panel}: The PBH mass function at formation time. \textit{Bottom-left panel}: The energy density of induced GW signal. \textit{Bottom-right panel}: The strain strength of induced GW signal.}
\label{fig:k4}
\end{figure}

In this Appendix, we display the smoother power spectrum employed to obtain Fig.~\ref{fig:GWstrain} of the main text. 
In particular, we start with a power spectrum that is displayed on the top left panel of Fig.~\ref{fig:k4}\footnote{While a smoothed power spectrum captures the most important features of the original delta-function power spectrum, we modify the peak location by a factor of $1.5$ and the peak amplitude by a factor of $3$ here, in order to reproduce a similar PBH mass function.}. The power spectrum is chosen to have peaks at $k = 10^{21}$ ${\rm Mpc}^{-1}$ corresponding to $f^{\rm peak}_{\rm GW} = 1.5~{\rm MHz} $. The corresponding PBH mass is $M_{\rm LPBH} = 10^4~{\rm g} $. The power spectrum is taken to  have a $k^4$ rise and fall for each side of the peak, following the results of \cite{Byrnes:2018txb}. We note that the results will not change qualitatively for other power-law behaviors.

The PBH mass function corresponding to our power spectrum is shown in the top right panel of Fig.~\ref{fig:k4}. On the bottom panels, we display the resulting GW spectrum. We note that in contrast to the GW spectrum depicted in Fig.~\ref{fig:GW2Peaks} resulting from the delta function power spectrum, the peaks in the GW spectrum are smoother.

\section{Models of Baryon Asymmetry Production from $X$ Decay}
\label{sec:susy}

There are many possibilities for baryogenesis model-building once the species $X$ has been produced by Hawking radiation. Here, we give two examples with a supersymmetric embedding. Both break $B-L$ symmetry. Since the DM in our scenario is comprised primarily of PBHs, the lightest neutralino  is not required to reproduce the observed relic density or even be stable. Although not the focus of our work, this can open up considerable freedom in the supersymmetric model-building side.

\subsection{Model 1: Multiple ($\geq 2$) Flavors of Gauge Singlets $X_\alpha$}

We  consider the Minimal Supersymmetric Standard Model (MSSM) extended with the following superfields: a single flavor of iso-singlet color triplets $N,~{\bar N}$ and at least two flavors of singlets $X_\alpha$. The superpotential of the theory is given by the MSSM superpotential along with
\be \label{superpot2}
W_{\rm extra} = \lambda_{i\alpha} X_{\alpha} u^c_i N + \lambda^\prime_{ij} d^c_id^c_j \overline{N} 
+ \frac{M_{\alpha}}{2} X_{\alpha} X_{\alpha} + M_{N} N \overline{N} \, .
\ee
The interference between tree-level and one-loop self-energy and vertex diagrams result in a baryon asymmetry (for $M_\alpha > M_N$). For  $X_\alpha \rightarrow N^* u^{c*}_i$ decay (with $X_\alpha$ being the fermionic component of the $X$ superfield), the asymmetry generated is
\be \label{Nasymmetry}
\gamma_{CP} = {\sum_{i,j,\beta} {\rm Im} \left(\lambda_{i \alpha} \lambda^*_{i \beta}\lambda^{*}_{j \beta}\lambda_{j\alpha}\right) \over 24 \pi \sum_{i}\lambda^{*}_{i\alpha} \lambda_{i\alpha}} ~ \left[3{\cal F}_S \left(M^2_\beta \over M^2_\alpha \right) + {\cal F}_V \left(M^2_\beta \over M^2_\alpha \right)\right] ,
\ee
where
\be
{\cal F}_S (x) = {2 \sqrt{x} \over x - 1} ~ ~ , ~ ~ {\cal F}_V = \sqrt{x} ~ {\rm ln} \left(1 + {1 \over x}\right) .
\ee
The extra factor of 3 in the denominator compared to standard expressions for the  asymmetry in leptogenesis appears due to the final state having baryon number $+1/3$. The factor 3 in the self energy contribution ${\cal F}_S$ comes from the sum over colors in intermediate states. The total asymmetry in this model can be obtained by summing up over contributions from scalar and fermionic components of all flavors of $X$, and the desired baryon asymmetry can be obtained by suitable choices of the couplings $\lambda_{i \alpha}$.

\subsection{Model 2: Multiple ($\geq 2$) Flavors of Iso-singlet Color Triplets $X_\alpha$}

A different variation is the case where the MSSM is extended by the following fields: $\geq 2$ flavors of iso-singlet color triplets $X_{\alpha},~\bar{X}_{\alpha}$ which have  hypercharges $+4/3,-4/3$ respectively; and a field $N$ that is a singlet under the SM gauge symmetry, but may be charged under a larger gauge group. The superpotential for the augmented sector is $W_{\rm extra}$, where
\be\label{superpot1}
W_{\rm extra} = \lambda_{i \alpha} N u^c_i X_{\alpha} + \lambda^\prime_{ij\alpha} d^c_i d^c_j
\overline{X}_{\alpha} 
+ {M_N \over 2} NN + M_{\alpha} X_{\alpha} \overline{X}_{\alpha}~.
\ee
Here $i,~j$ denote MSSM flavor indices with color indices suppressed. The couplings  $\lambda^\prime_{ij\alpha}$ are antisymmetric under $i \leftrightarrow j$. The fermionic components of $N$ and $X,~{\bar X}$ are assigned charges $+1$ and $-1$, respectively,  under $R$-parity. While this ensures $R$-parity conservation, this is not strictly required since we do not demand that the lightest supersymmetric particle constitutes the majority of DM. We denote by $X_\alpha,~\psi_\alpha,~{\bar X}_\alpha,~{\bar \psi}_\alpha$ the scalar and fermionic components of the superfields $X_\alpha$. Scalar components of a given flavor have mass eigenvalues given by
\be \label{mass}
m^2_\alpha = |M_{\alpha}|^2 + \tilde{m}_{\alpha}^2 \pm \vert B_\alpha M_{\alpha} \vert ,
\ee
where ${\tilde m}_\alpha$ is the soft mass of scalars and $B_\alpha$ is the $B$-term associated with the superpotential mass term $M_\alpha X_\alpha {\bar X}_\alpha$.

The baryon asymmetry obtained in this model can be obtained as follows. For simplicity, we consider two flavors of color triplets, which is the minimum number required to obtain an asymmetry, and focus on decays with supersymmetry conserving interactions. For the fermionic component $\psi_1$, the relevant decay modes (if kinematically allowed) are ${\bar \psi}_1 \rightarrow d^{c*}_i {\tilde d}^{c*}_j$, with $\Delta B = +2/3$, and ${\bar \psi}_1 \rightarrow {\tilde N} u^c_k,~N {\tilde u}^c_i$ for which $\Delta B = -1/3$. Here $N,{\tilde N}$ are the fermionic and scalar components of the superfield $N$. The interference between the tree-level decay diagram and the one-loop self-energy diagram results in a baryon asymmetry from ${\bar \psi}_1$ and ${\bar \psi}^*_1$ decays (if $M_1 > M_N$) as follows:
\be \label{asymmetry}
\epsilon_1 = {1 \over 8 \pi} ~ {\sum_{i,j,k} {\rm Im} \left(\lambda^*_{k1}\lambda_{k2}\lambda^{\prime *}_{ij1}\lambda^{\prime}_{ij2}\right) \over \sum_{i,j}\lambda^{\prime *}_{ij1} \lambda^{\prime}_{ij1} + \sum_{k}\lambda^*_{k1} \lambda_{k1}} ~ {\cal F}_S \left(M^2_2 \over M^2_1 \right), 
\ee
where, for $M_2 - M_1 > \Gamma_{{\bar \psi}_1}$, 
\be \label{self}
{\cal F}_S(x) = {2 \sqrt{x} \over x - 1}.
\ee
This is similar to the result in  standard leptogenesis~\cite{Barrow:2022gsu} with the difference that in this model there are no vertex diagrams. The same asymmetry is obtained from ${\psi}_1$ and $\psi^*_1$ decays, since ${\bar \psi}_1$ and $\psi^c_1$ constitute a four-component spinor with hypercharge  $-4/3$.  The same asymmetry is also obtained from the decay of scalars $X_1,~{\bar X}_1$ and their antiparticles $X^*_1,~{\bar X}^*_1$, in the limit of unbroken supersymmetry; when supersymmetry is broken the asymmetries from scalar and fermion  decays are similar if their masses are similar,  $m_{1,2} \sim M_{1,2}$. For $M_2 > M_N$ scalar and fermionic components of $X_2,{\bar X}_2$ will decay to produce an asymmetry $\epsilon_2$, which can be obtained from the expression for $\epsilon_1$ with $1 \leftrightarrow 2$. The total asymmetry in this model can be obtained as $\epsilon_1 + \epsilon_2$.

\bibliography{draft}
\bibliographystyle{JHEP}

\end{document}